\documentclass[letterpaper, 11pt]{article}

\usepackage[nocompress]{cite}
\usepackage{jheppub}

\usepackage{graphicx}
\usepackage{epstopdf}
\usepackage{amsmath, amssymb}

\usepackage{bm}
\usepackage{bbold}
\usepackage{caption}
\usepackage{subcaption}

\usepackage{multirow} 
\usepackage{longtable} 

\usepackage{romannum}

\newcommand{\be}{\begin{eqnarray}}
\newcommand{\ee}{\end{eqnarray}}
\newcommand{\nn}{\nonumber}
\newcommand{\bn}{\begin{enumerate}}
\newcommand{\en}{\end{enumerate}}

\def\CF{{\cal F}}

\def\CI{{\cal I}}

\def\CN{{\cal N}}
\def\CO{{\cal O}}

\def\CT{{\cal T}}

\def\a{\alpha}

\def\e{\epsilon}

\def\k{\kappa}

\def\s{\sigma}

\def\G{\Gamma}

\def\half{\frac{1}{2}}

\def\vev#1{\langle #1 \rangle}

\def\Tr{{\rm Tr}}
\def\tr{{\rm Tr}}

\def\vec#1{\bm{#1}}

\newcommand{\bea}{\begin{eqnarray}}
\newcommand{\eea}{\end{eqnarray}}

\def\CF{{\cal F}}

\def\CI{{\cal I}}

\def\CN{{\cal N}}
\def\CO{{\cal O}}

\def\CT{{\cal T}}

\def\a{\alpha}

\def\e{\epsilon}

\def\k{\kappa}

\def\s{\sigma}

\def\G{\Gamma}

\def\half{\frac{1}{2}}

\def\vev#1{\langle #1 \rangle}

\def\Tr{{\rm Tr}}
\def\tr{{\rm Tr}}

\title{$\CN=1$ Deformations and RG Flows of $\CN=2$ SCFTs, Part \Romannum{2}: Non-principal deformations}

\author[a]{Prarit Agarwal,}
\author[b]{Kazunobu Maruyoshi,}
\author[c]{and Jaewon Song}
\affiliation[a]{Department of Physics and Astronomy \& Center for Theoretical Physics\\ Seoul National University, Seoul 151-747, Korea}
\affiliation[b]{Faculty of Science and Technology, Seikei University\\ 3-3-1 Kichijoji-Kitamachi, Musashino-shi, Tokyo, 180-8633, Japan}
\affiliation[c]{Department of Physics, University of California, San Diego \\La Jolla, CA 92093, USA}
\emailAdd{agarwalprarit@gmail.com}
\emailAdd{maruyoshi@st.seikei.ac.jp}
\emailAdd{jsong@physics.ucsd.edu}

\abstract
{
We continue to investigate the $\CN=1$ deformations of four-dimensional $\CN=2$ superconformal field theories (SCFTs) labeled by a nilpotent element of the flavor symmetry \cite{Maruyoshi:2016aim}. This triggers a renormalization group (RG) flow to an $\CN=1$ SCFT. We systematically analyze all possible deformations of this type for certain classes of $\CN=2$ SCFTs: conformal SQCDs, generalized Argyres-Douglas theories and the $E_6$ SCFT. We find a number of examples where the amount of supersymmetry gets enhanced to $\CN=2$ at the end point of the RG flow. Most notably, we find that the $SU(N)$ and $Sp(N)$ conformal SQCDs can be deformed to flow to the Argyres-Douglas (AD) theories of type $(A_1, D_{2N-1})$ and $(A_1, D_{2N})$ respectively. This RG flow therefore allows us to compute the full superconformal index of the $(A_1,D_N)$ class of AD theories. Moreover, we find an infrared duality between $\CN=1$ theories where the fixed point is described by an $\CN=2$ AD theory. 
We observe that the classes of examples that exhibit supersymmetry enhancement saturate certain bounds for the central charges implied by the associated two-dimensional chiral algebra.
 
}

\preprint{SNUTP16-006}

\begin{document}
\maketitle

\section{Introduction}

The study of quantum field theories in the strongly coupled regime is notoriously challenging owing to the inapplicability of perturbative analysis.  However, this problem becomes somewhat tractable when one focuses on supersymmetric field theories.  This is largely because the quantum corrections to various physical quantities of interest are strongly constrained by holomorphy \cite{Seiberg:1994bp}.  This has made many exact computations possible, as a result of which supersymmetric theories have become a testing ground for many novel approaches being developed to study quantum field theories.

In this paper, we study the renormalization group (RG) flow of certain four-dimensional $\CN=1$ supersymmetric field theories obtained by considering a specific category of $\CN=1$ preserving deformations \cite{Maruyoshi:2016aim} of $\CN=2$ superconformal field theories (SCFTs) with non-Abelian flavor symmetry $F$.  These correspond to coupling a gauge-singlet field, $M$, to the moment map operator, $\mu$ (which is the scalar component in the $\CN=2$ supermultiplet of the conserved flavor current) via
\begin{align}
 W = \tr M \mu \ , 
\end{align}
and then giving a nilpotent vev to $M$. A nilpotent element in $F$ is classified by its $SU(2)$ embeddings $\rho: SU(2) \hookrightarrow F$ and given by $\rho(\s^+)$. Therefore one can obtain an $\CN=1$ SCFT $\CT_{{\rm IR}}$ labelled by an $\CN=2$ SCFT $\CT_{{\rm UV}}$ and the $SU(2)$ embedding $\rho$ as:
\begin{align}
 \CT_{{\rm UV}} \rightsquigarrow \CT_{{\rm IR}}[\CT_{{\rm UV}}, \rho].
\end{align}
Deformations of this kind were previously considered in \cite{Gadde:2013fma,Agarwal:2013uga,Agarwal:2014rua,Agarwal:2015vla,Fazzi:2016eec}.

In \cite{Maruyoshi:2016tqk, Maruyoshi:2016aim}, the last two authors of the current paper demonstrated that the four-dimensional $\CN=1$ theories obtained in this manner have rich dynamics characterized by operator decoupling and appearance of accidental symmetries along their RG flow. The main tool of analysis for such RG flows was the principle of $a$-maximization \cite{Intriligator:2003jj} and its modification \cite{Kutasov:2003iy}.  What was perhaps most surprising is the fact that many of these theories flow to IR fixed points at which there is an enhancement of supersymmetry from $\CN=1$ to $\CN=2$. 
By investigating the RG flows, certain $\CN=1$ Lagrangians were discovered, whose IR fixed points were found to be the Argyres-Douglas theory \cite{Argyres:1995jj} and its generalization of $(A_1, A_N)$ type.  This made it possible to obtain the full superconformal indices of the these theories. 

However, the deformation analyzed in \cite{Maruyoshi:2016tqk, Maruyoshi:2016aim} belonged to only one particular case among many choices. It was obtained by giving $M$, the nilpotent vev corresponding to the principal (maximal) nilpotent orbit of the flavor symmetry $F$ of the undeformed $\CN=2$ theory, which breaks $F$ completely. An immediate question that arises in this context, is if the above mentioned phenomenon continues to be true when the vev of $M$ is given by other nilpotent orbits of $F$. It is this question that we seek to answer in the current paper. We will show that indeed there exists a class of nilpotent vevs that is different from the principal case and yet triggers an RG flow to an IR fixed point with the enhanced supersymmetry. In particular, this will enable us to write $\CN=1$ Lagrangians flowing to the so-called $(A_1,D_N)$ theories, thereby allowing us to compute their full superconformal indices.

The Argyres-Douglas theory and its generalizations are believed to be some of the simplest known $\CN=2$ SCFTs. They are characterized by the fact that their Coulomb branch operators have fractional scaling dimensions. They were originally found at 
special loci on the Coulomb branches of $\CN=2$ supersymmetric gauge theories \cite{Argyres:1995jj, Argyres:1995xn} (also see \cite{Eguchi:1996vu, Eguchi:1996ds} for many more examples) where the massless spectra  consist of particles with mutually non-local electromagnetic charges.  The lack of a duality frame in which all the particles are only electrically charged then makes it impossible to write an $\CN=2$ Lagrangian describing this system, hence giving rise to the belief that these theories are isolated strongly coupled SCFTs. A more modern approach towards constructing Argyres-Douglas theories consists of wrapping M5-branes on a sphere with one irregular puncture and at most one regular puncture \cite{Gaiotto:2009hg,Bonelli:2011aa,Gaiotto:2012sf, Xie:2012hs}. 

The lack of a Lagrangian makes it difficult to compute any physically relevant data. 
Progress in our understanding of theses theories might have been slow, but has not been completely stunted. Holographic techniques were successfully employed in \cite{Aharony:2007dj} to compute the central charges for the so-called $H_0$, $H_1$, and $H_2$ theories. This result was confirmed by a field theoretic method in \cite{Shapere:2008zf}. This technique was later applied to the generalized AD theories in \cite{Xie:2013jc}. 
Their BPS particle spectrum in the Coulomb branch was carefully studied in \cite{Shapere:1999xr,Gaiotto:2009hg,Cecotti:2010fi,Cecotti:2011rv, Alim:2011ae,Alim:2011kw, Maruyoshi:2013fwa}. The two-dimensional chiral algebra (in the sense of \cite{Beem:2013sza}) for AD theories corresponds to non-unitary minimal models. Moreover, some of the AD theories saturate the lower bound of the conformal anomaly $c$ and the flavor central charge $k$ \cite{Liendo:2015ofa, Lemos:2015orc}. 
The authors of \cite{Cordova:2015nma} found a relation between the Schur limit of the superconformal index and the BPS particle spectrum, building upon the results in \cite{Cecotti:2010fi,Iqbal:2012xm}. This relation was further developed in \cite{Cecotti:2015lab,Cordova:2016uwk}. Using this, they computed the Schur indices of the generalized AD theories and found that it is identical to the vacuum character of the two-dimensional chiral algebra.  The Schur, Macdonald and Hall-Littlewood indices were independently obtained in \cite{Buican:2015ina,Buican:2015tda,Song:2015wta} where the authors were able to take advantage of the 2d/4d correspondence proposed in \cite{Gadde:2009kb,Gadde:2010te,Gadde:2011ik,Gadde:2011uv,Gaiotto:2012xa}. We will use these limiting cases to provide a non-trivial check of our proposal for the full superconformal indices of the SCFTs of type $(A_1,D_N)$.

\paragraph{Summary of results}
Through our analysis, we find that when the undeformed $\CN=2$ SCFT is given by an $SU(N)$ gauge theory with $N_f=2N$ fundamental hypermultiplets and $M$ is given a vev corresponding to the 
next-to-principal nilpotent orbit labeled by the partition $[N-1,1]$ of the $SU(2N)$ flavor symmetry, the IR fixed point is characterized by the $(A_1,D_{2N})$ theory. The vev of $M$ preserves a $U(1)^2 (\subset SU(2N) \times U(1))$ flavor symmetry. We claim that this enhances to the $SU(2) \times U(1)$ flavor symmetry of the corresponding theory. This claim is supported by the fact that upon appropriate normalization of the $U(1)^2$ fugacities, they arrange themselves into $SU(2)\times U(1)$ characters in the superconformal index. Similarly, when the undeformed theory is an $\CN=2$ $Sp(N)$ gauge theory with $4N+4$ fundamental half-hypermultiplets and $M$ is given a vev corresponding to the nilpotent orbit of $SO(4N+4)$ labelled by the partition $[4N+1,1^3]$, the IR fixed point corresponds to the $(A_1,D_{2N+1})$ theory. This vev of $M$ preserves an $SO(3) (\subset SO(4N+4))$ which is isomorphic to the $SU(2)$ flavor symmetry of the corresponding AD theories in the IR. 
\begin{table}
\centering
\begin{tabular}{|c|c|c|}
	\hline
	$\CT_{UV}$ & $\rho$ & $\CT_{IR}[\CT_{UV}, \rho]$ \\
	\hline \hline
	\multirow{2}{*}{$SU(N)$ with $N_f=2N$} &$[N]$  & $(A_1, A_{2N-1})$ theory \\
	& $[N-1, 1]$  & $(A_1, D_{2N})$ theory \\
	\hline
	\multirow{2}{*}{$Sp(N)$ with $N_f=2N+2$} &$[4N+4]$  & $(A_1, A_{2N})$ theory \\
	  & $[4N+1, 1^3]$  & $(A_1, D_{2N+1})$ theory \\
	\hline
	$(I_{N, k}, F)$ & $[N]$ & $(A_{N-1}, A_{N+k-1})$ theory \\
	$(I_{N, -N+2}, F)$ & $[N-1, 1]$ & $(A_1, D_N)$ theory \\
	\hline
	\multirow{3}{*}{$E_6$ SCFT} & $E_6$ & $H_0$ theory \\
	& $D_5$ & $H_1$ theory \\
	& $D_4$ & $H_2$ theory \\
	\hline
\end{tabular}
\caption{Summary of results: Here we list the RG flows exhibiting supersymmetry enhancement. }
\label{table:summary}
\end{table}
  These, in addition to the cases found in \cite{Maruyoshi:2016aim} for the principal embedding, are summarized in table \ref{table:summary}.
 
We also consider the effect of giving $M$ a vev corresponding to other nilpotent orbits of the respective flavor groups. However, in these cases, the IR theory does not seem to exhibit any supersymmetry enhancement. For most of these cases the central charges of the IR fixed point are irrational and hence cannot posses an $\CN=2$ supersymmetry \cite{Maruyoshi:2016aim}. However, there are few cases, other than those mentioned above, where the IR central charges do become rational. We were not able to find any $\CN=2$ SCFTs to which they might correspond.  Neither were we able to find any particular pattern governing the partitions for which the central charges are rational. For the sake of completeness, in tables \ref{tab:RationalChargesSU} and \ref{tab:RationalChargesSp},  we list the respective partitions of $SU(2N)$ and $SO(4N+4)$ for which central charges are rational.

Similar deformation of the $\CN=2$ $SO(N)$ gauge theory coupled to $2N-4$ fundamental half-hypermultiplets with $F=Sp(N-2)$ can also be considered. In this case, the vev corresponding to the principal nilpotent orbit always seems to give irrational central charges and so the IR theory cannot be invariant under an $\CN=2$ supersymmetry. Other orbits for which the central charges become rational pop-up at apparently random places as we scan through the various values of $N$. We list these in table \ref{tab:RationalChargesSO}. Once again, we were not able to find any $\CN=2$ theories that might be associated to their IR fixed points. 

Moreover, we consider deformations of the generalized AD theory of type $(I_{N, k}, F)$ \cite{Xie:2013jc}, which has (at least) $SU(N)$ flavor symmetry. We find that the deformation corresponding to the principal embedding triggers a flow to the AD theory of type $(A_{N-1}, A_{N+k-1})$. When $k=-N+2$, the non-principal embedding $[N-1, 1]$ gives the $(A_1, D_N)$ theory. 
When $N=2n$ is even, $(I_{2n, -2n+2}, F)$ theory is identical to the $SU(n)$ conformal SQCD, which gets back to the previous analysis. Thus it is interesting for $N=2n+1$ odd. 
\begin{figure}[!h]
\begin{align} \nn
\begin{array}{ccc}
\boxed{Sp(n) \textrm{ SQCD with } N_f=2n+2} & \leftrightarrow &
\boxed{ (I_{2n+1, -2n+1}, F) \textrm{ AD theory}}\quad \\
\quad \rho=[4n+1, 1^3] \searrow & & \swarrow \rho=[2n, 1] \qquad \quad\\
\multicolumn{3}{c}{ \boxed{(A_1, D_{2n+1}) \textrm{ AD theory} }}
\end{array}
\end{align}
\caption{IR duality among $\CN=1$ theories. The IR fixed point is described by $\CN=2$ AD theory}
\end{figure}
When combined with the RG flow from the deformed $Sp(n)$ conformal SQCD to $(A_1, D_{2n+1})$ theory, this example provides us with a novel IR duality, where two distinct $\CN=1$ theories flow to the same IR fixed point theory with the $\CN=2$ supersymmetry. 

Deformations of the $E_6$ SCFT of \cite{Minahan:1996fg} are also considered. In this case we find three deformations with RG flows to $\CN=2$ fixed points.

\paragraph{Chiral algebra and SUSY enhancement} 
While we are not aware of the mechanism of the supersymmetry enhancement, 
it is worthwhile to point out that the condition of the enhancement is somewhat related to the property of the associated two-dimensional chiral algebra \cite{Beem:2013sza} $\chi[\CT_{{\rm UV}}]$ of the undeformed theory $\CT_{{\rm UV}}$.
We conjecture that the IR theory experiences the supersymmetry enhancement for the following two cases:
  \begin{enumerate}
  \item $\chi[\CT_{{\rm UV}}]$ is given by the affine Kac-Moody algebra $\hat{\mathfrak{f}}$ (affine version of the flavor symmetry group $F$) where the 2d stress tensor is given by the Sugawara construction, and the deformation $\rho$ corresponds to the maximal (principal) nilpotent orbit of $F$.
  \item $\chi[\CT_{{\rm UV}}]$ is given by the affine Kac-Moody algebra $\hat{\mathfrak{f}}$ as above. In addition,  the flavor central charge $k_F$ saturates the bound given in table \ref{tab:bound}, and the deformation $\rho$ corresponds to the next-to-maximal nilpotent orbit which preserves some amount of $F$. 
  \end{enumerate}
When the 2d stress tensor is given by the Sugawara construction, the central charges saturates the bound, in terms of the four-dimensional central charges, given as
  \bea
  \frac{\dim F}{c} \geq \frac{24 h^\vee}{k_F} -12, 
  \label{sugawara}
  \eea
where $h^\vee$ is the dual Coxeter number of $F$.
The flavor central charge bound is shown in table \ref{tab:bound}.
  \begin{table}
\centering
\begin{tabular}{|cc|c|}
\hline
	$F$ & & $k_F$ bound \\
	\hline \hline
	$SU(2)$ & $\cdot$ & $\frac{8}{3}$ \\
	$SU(N)$ & $N \geq 3$ & $N$ \\
	$Sp(N)$ & $N \geq 3$ &  $N+2$\\
	$SO(N)$ & $N = 4,\ldots,8$ & $4$\\
	$SO(N)$ & $N \geq 8$ & $N-4$\\
	\hline
\end{tabular}
\caption{The lower bound for the flavor central charge $k_F$ for $F=SU(N)$, $Sp(N)$ and $SO(N)$ \cite{Beem:2013sza, Lemos:2015orc}.} 
\label{tab:bound}
\end{table}
 The nilpotent orbit is maximal if its dimension is the highest one. 
The next-to-maximal is meant in this sense. For the $\mathfrak{su}(n)$, the next-to-maximal or the subregular orbit is given by the partition $[n-1, 1]$, which preserves $\mathfrak{u}(1)$ subgroup. 
  Note that in the $\mathfrak{so}(2k)$ case, the next-to-maximal (subregular) orbit does not preserve any flavor symmetry. 
  Instead, the one with the highest dimension and preserving some of the flavor symmetry is $[2k-3,1^3]$. For the $\mathfrak{e}_6$, the `next-to-maximal' with some unbroken flavor symmetry would be $D_5$ in terms of the Bala-Carter label, not the subregular orbit $E_6(a_1)$.  
  
This conjecture is indeed true for all the examples we consider in this paper. It would be interesting to find a proof or an explanation behind this phenomena. 
  


\paragraph{Organization}
The organization of this paper is as follows.
In section \ref{sec:deformation}, we review the procedure of our class of $\CN=1$ deformations of $\CN=2$ SCFTs. 
We then give the formulae for the 't Hooft anomaly coefficients of the $R$-symmetries in the cases when the flavor symmetry of the original $\CN=2$ SCFT is $SU(N)$, $Sp(N)$ and $SO(N)$.
In section \ref{sec:SQCD}, we analyze the RG flow triggered by the deformation corresponding to the next-to-principal nilpotent orbit in the case of
$\CN=2$ conformal SQCD with classical gauge group $SU(N)$, $Sp(N)$ and $SO(N)$.
In section \ref{sec:genAD}, we consider the deformations of the generalized AD theory and the $E_6$ SCFT. 
In section \ref{sec:SCIA1DN}, we compute the full superconformal indices of the $(A_1, D_N)$ theories using the ``Lagrangian descriptions" we find in section \ref{sec:SQCD}. We also use this result to compute the indices of $(A_3, A_3)$ and $(A_2, A_5)$ theories and check their invariance under S-dualities. 
In the appendix, we discuss the superconformal index of adjoint SQCD in the scenario when some of the operators decouple along the RG flow. 

\section{$\CN=1$ deformations of $\CN=2$ SCFTs with flavor symmetry $F$}
\label{sec:deformation}
  In this section we consider the $\CN=1$ deformation procedure introduced in \cite{Maruyoshi:2016aim}
  which is applicable to arbitrary $\CN=2$ SCFT $\CT_{{\rm UV}}$ with a non-Abelian flavor symmetry $F$
  (here $F$ could be a subgroup of the full flavor symmetry of $\CT_{{\rm UV}}$).
  Let us denote the Lie algebra of $F$ as $\mathfrak{f}$.
  To this SCFT,
    \begin{itemize}
    \item we couple an $\CN=1$ chiral multiplet $M$ transforming in the adjoint representation of $\mathfrak{f}$ with the superpontential
      \bea
      W
       =     \tr M \mu,
       \label{superpotential}
      \eea
      where $\mu$ is the moment map operator which is the lowest component of the conserved current multiplet of $F$, and
    \item give a nilpotent vev to $M$: $\langle M \rangle =\rho(\sigma^+)$, 
             where $\rho$ is the embedding $\rho$: $\mathfrak{su}(2) \rightarrow \mathfrak{f}$.
    \end{itemize}
  
  Let the generators of the Cartan of $SU(2)_R$ and $U(1)_r$ symmetries be $I_3$ and $r$ respectively\footnote{Here we use the convention that the $\mu$ operator has $2I_3$ charge $2$, and Coulomb branch operators have dimensions $\Delta = \frac{r}{2}$.},
  and we denote them as
    \bea
    (J_+, J_-)
     =     (2 I_3, r).
    \eea
  In this convention, the charges of $M$ and $\mu$ are $(J_+, J_-) = (0,2)$ and $(2,0)$ respectively.
  The vev of $M$ breaks $U(1)_{J_-}$ symmetry, but the following combination is preserved:
    \bea
    J_- - 2 \rho(\sigma_3) 
    \label{shift}
    \eea
  We now decompose $M$ and $\mu$ into irreducible representations of $SU(2) \subset F$, in accordance with decomposition of the adjoint representation of $\mathfrak{f}$, ${\bf adj} \rightarrow \oplus_j V_j$, where $V_j$ is spin-$j$ representation of $SU(2)$.
  
 
     As studied in \cite{Gadde:2013fma, Agarwal:2013uga, Agarwal:2014rua, Agarwal:2015vla, Maruyoshi:2016aim}, for each spin-$j$ representation of $M$, only the component with $j_3 = -j$,  will stay coupled to the theory.    
  The superpotential thus becomes
    \bea
    W
     =     \sum_j M_{j, -j} \mu_{j, j},
     \label{finalW}
    \eea
  where $M_{j,-j}$ has charge $ (J_+, J_-) = (0, 2+2j)$.
  
  Let us now give the formulas of the anomaly coefficients.
  Henceforth, we will denote the the central charges $a$ and $c$ of $\CT_{{\rm UV}}$ as $a_\CT$ and $c_\CT$ respectively.
  In terms of these, the anomalies of $\CT_{{\rm UV}}$ are given by
    \begin{align}
    \begin{split}
    \tr J_+
    =     \tr J_+^3
    &=    0, \\
    \tr J_-
     =    \tr J_-^3
    &=   48(a_{\CT}-c_{\CT}), \\
    \tr J_+^2 J_- 
    &=   8 (2a_{\CT} - c_{\CT}),  \\
    \tr J_+ J_-^2
    &=    0,  \\
    \tr J_- T^a T^a
    &= - \frac{k_F}{2},
    \end{split}
    \label{N=2anom}
    \end{align}
  where $T_a$ are the generators of $\mathfrak{f}$ and $k_F$ is the flavor central charge.
  After the deformation, we will have to account for the shift \eqref{shift} and the contribution from the remaining $M$ multiplets \eqref{finalW}. 
  The former only changes the $\Tr J_-^3$ anomaly as $\Tr J_-^3 \rightarrow \Tr J_-^3 + 12 \Tr J_- \rho(\sigma_3)^2 = \Tr J_-^3 -6 k_F I_Y$,
  where $I_Y$ is the embedding index.
  The latter contribution can be easily computed once we consider the decomposition of the adjoint representation of $\mathfrak{f}$. 
  
  Let us now see the explicit formulas of the anomaly coefficients for $F=SU, Sp$ and $SO$.

\paragraph{$F = SU(N)$ case}
  The embedding $\rho$ is specified by a partition of $N$ (or a Young diagram with $N$ boxes).
  We denote this by $N = \sum_k k n_k$.
  Due to the vev to $M$, the flavor symmetry $F$ is broken to $S [\prod_k U(n_k)]$, where $S[\ldots]$ means the traceless part of $\ldots$.
  
  In this case the embedding index is given by 
    \bea
    I_{Y} = \frac{1}{6} \sum_{k=1}^\ell k (k^2-1) n_k.
    \eea
  The components of $M$ transform in representations of the remaining global symmetry.  
  By adding these contributions one gets the anomalies of the deformed theory:
    \bea
    \tr J_+
    &=&    \tr J_+^3
     =   - \sum_{k=1}^\ell k n_k^2 - 2 \sum_{k<l} k n_k n_l + 1,
              \nn \\
    \tr J_-
    &=&  48(a_{\CT}-c_{\CT}) + N^2 - 1,
             \nonumber \\
    \tr J_-^3
    &=&   48(a_{\CT}-c_{\CT}) - 6k_F I_Y + \sum_{k=1}^\ell k^2(2k^2-1) n_k^2 + 2 \sum_{k<l} k l(k^2 + l^2 -1) n_k n_l - 1,
             \nn \\
    \tr J_+^2 J_-
    &=&    8 (2a_{\CT} - c_{\CT}) + N^2  - 1,
           \nn \\
    \tr J_+ J_-^2
    &=& - \sum_{k=1}^\ell \frac{k(4k^2-1)}{3} n_k^2 - 2 \sum_{k<l} \frac{k(3l^2 + k^2 - 1)}{3} n_k n_l + 1,
    \label{totalanom}
    \eea
  where we have used the identity $N^2 =  \sum_{k=1}^\ell k^2 n_k^2 + 2 \sum_{k<l} k l n_k n_l $.

\paragraph{$F=SO(N)$ case}
In the case of the global symmetry being $SO(N)$, the embeddings $\rho$ are in one-to-one correspondence with those partitions of $2N$, for which the even parts occur with even multiplicity\footnote{As mentioned in \cite{collingwood17008nilpotent}, when $N$ is even, an exception to this rule comes from partitions which consists of only even parts, each appearing with even multiplicity. Such partitions are called ``very even'' and correspond to two distinct embeddings, which are exchanged under the action of the $\mathbb{Z}_2$ outer-automorphism of $SO(N)|_{N=even}$. This distinction between the two embeddings associated to ``very even'' partitions was important in \cite{Chacaltana:2011ze, Chacaltana:2013oka, Lemos:2012ph}, but seems to be inconsequential for the deformations studied here. We will therefore treat the two embeddings corresponding to any given ``very even'' partition, to be equivalent.}. A generic partition specifying an embedding is therefore given by 
\be
N=\sum_{k_e} k_e n_{k_e} + \sum_{k_o} k_o n_{k_o} \ , \text{$\forall k_e$, $n_{k_e} =$ even}  \ .
\ee 
The corresponding vev for $M$ breaks the $SO(N)$ symmetry down to $\prod_{k_o} SO(n_{k_o}) \times \prod_{k_e} Sp(\frac{n_{k_e}}{2})$. 
The embedding index is given by \cite{Panyushev201515}
\be
I_Y= \frac{1}{12} \sum_{k_e}k_e (k_e^2-1) n_{k_e}  +  \frac{1}{12} \sum_{k_o}k_o (k_o^2-1) n_{k_o} \ ,
\label{embeddingso}
\ee
where we have normalized the $SO(N)$ generators to be such that the quadratic index, $\Tr T^a T^b$, of the fundamental representation is 1.

Upon decomposing $M$ into representations of $SU(2) \times \prod_{k_o} SO(n_{k_o}) \times \prod_{k_e} Sp(\frac{n_{k_e}}{2})$ and keeping only the $j_3 = -j$ component for each spin-$j$ representation of $M$, the anomalies of the deformed theory are given by
 \begin{align}
 \tr J_+
 &=   \tr J_+^3
 =   - \half \sum_{k_e} k_e n_{k_e}^2 - \frac{1}{2} \sum_{k_o} (k_o n_{k_o}-1) n_{k_o}  \nonumber \\
 &~~ ~~~~~ - \sum_{k_e<l_e} k_e n_{k_e} n_{l_e} - \sum_{k_o<l_o} k_o n_{k_o} n_{l_o} - \sum_{k_e,k_o} min(k_e,k_o) n_{k_e} n_{k_o} \ , \nonumber \\    
 \tr J_-
 &=  48(a_{\CT}-c_{\CT}) + \half N(N-1),
 \nonumber \\
 \tr J_-^3
 &=   48(a_{\CT}-c_{\CT}) - 6k_F I_Y + \half \sum_{k_o} k_o n_{k_o} (2k_o^3 n_{k_o} - 4k_o^2 - k_o n_{k_o}+3) \nonumber \\ 
 &  + \half \sum_{k_e} k_e n_{k_e} (2k_e^3 n_{k_e} - 4k_e^2 - k_e n_{k_e}+3) +  \sum_{k_e<l_e} k_e l_e (k_e^2+l_e^2-1) n_{k_e} n_{l_e}  \nonumber \\  
 & +  \sum_{k_o<l_o} k_o l_o(k_o^2+l_o^2-1) n_{k_o} n_{l_o} + 
  \sum_{k_o, k_e} k_o k_e (k_o^2+k_e^2-1)  n_{k_o} n_{k_e} \ ,
 \nn \\
 \tr J_+^2 J_-
 &=    8 (2a_{\CT}-c_{\CT}) + \half N(N-1),
 \nonumber \\
 \tr J_+ J_-^2
 &= -\frac{1}{6}\sum_{k_o} n_{k_o} \left(4 k_o^3 n_{k_o}-6 k_o^2-k_o n_{k_o}+3\right) -\frac{1}{6}\sum_{k_e} k_e n_{k_e} ( 4 k_e^2 n_{k_e} - 6 k_e - n_{k_e}) \nn \\
 & - \frac{1}{3}\sum_{k_e < l_e} k_e \left(3 l_e^2+k_e^2-1\right)n_{k_e} n_{l_e} -\frac{1}{3}\sum_{k_o < l_o} k_o \left(3 l_o^2+k_o^2-1\right)n_{k_o} n_{l_o} \nn \\
 & - \frac{1}{3}\sum_{k_e < k_o} k_e \left(3 k_o^2+k_e^2-1\right)n_{k_e} n_{k_o} -\frac{1}{3}\sum_{k_o < k_e} k_o \left(3 k_e^2+k_o^2-1\right)n_{k_o} n_{k_e} \ .
 \label{eq:totalanomSO}
 \end{align}

\paragraph{$F=Sp(N)$ case}
When the flavor symmetry is given by the rank-$N$ symplectic group, $Sp(N)$, the embeddings $\rho$ are in one-to-one correspondence with those partitions of $2N$ for which the odd parts occur with even multiplicity \cite{collingwood17008nilpotent}. Let us write this as 
\be
2N=\sum_{k_e} k_e n_{k_e} + \sum_{k_o} k_o n_{k_o} \ , \text{$\forall k_o$, $n_{k_o} =$ even}  \ ,
\ee
where $k_e$ and $k_o$ denote even and odd parts of the partition respectively. The corresponding vev for $M$ breaks the flavor symmetry down to $\prod_{k_e} SO(n_{k_e}) \times \prod_{k_o} Sp(\half n_{k_o})$. The embedding index is given by the same formula \eqref{embeddingso},
where the $Sp(N)$ generators are normalized to be such that the quadratic index, $\Tr T^a T^b$, of the fundamental representation is 1.

We can now decompose $M$ into representations of $SU(2) \times \prod_{k_e} SO(n_{k_e}) \times \prod_{k_o} Sp(\half n_{k_o})$, where $SU(2)$ corresponds to the embedding, $\rho$. Recall from \eqref{finalW}, for each spin-$j$ representation of $M$, only the components with $j_3 = -j$ will survive in the IR. Once the contribution of these components are taken into account, the anomalies of the deformed theory are given by
 \begin{align}
    \tr J_+
    &=    \tr J_+^3
     =   - \half \sum_{k_e} k_e n_{k_e}^2 - \frac{1}{2} \sum_{k_o} (k_o n_{k_o}+1) n_{k_o}   \nonumber \\
      &~~~~~~~ - \sum_{k_e<l_e} k_e n_{k_e} n_{l_e} - \sum_{k_o<l_o} k_o n_{k_o} n_{l_o} - \sum_{k_e,k_o} min(k_e,k_o) n_{k_e} n_{k_o} \ , \nonumber \\    
   \tr J_-
   &=  48(a_\CT-c_\CT) + 2N^2 + N,
             \nonumber \\
    \tr J_-^3
    &=   48(a_\CT-c_\CT) - 6k_F I_Y + \half \sum_{k_o} k_o n_{k_o} (2k_o^3 n_{k_o} + 4k_o^2 - k_o n_{k_o}-3)  \nonumber \\ 
     & + \half \sum_{k_e} k_e n_{k_e} (2k_e^3 n_{k_e} + 4k_e^2 - k_e n_{k_e}-3) +\sum_{k_e<l_e} k_e l_e (k_e^2+l_e^2-1) n_{k_e} n_{l_e}   \nonumber \\  
     &  +  \sum_{k_o<l_o} k_o l_o(k_o^2+l_o^2-1) n_{k_o} n_{l_o} + \sum_{k_o, k_e} k_o k_e (k_o^2+k_e^2-1)  n_{k_o} n_{k_e}, \nn  \\ 
    \tr J_+^2 J_-
    &=    8 (2a_\CT-c_\CT) + 2N^2 + N,
    \nonumber \\
    \tr J_+ J_-^2
    &= -\frac{1}{6}\sum_{k_o} n_{k_o} \left(4 k_o^3 n_{k_o}+6 k_o^2-k_o n_{k_o}-3\right) -\frac{1}{6}\sum_{k_e} k_e n_{k_e} ( 4 k_e^2 n_{k_e} + 6 k_e - n_{k_e}) \nn \\
    & - \frac{1}{3}\sum_{k_e < l_e} k_e \left(3 l_e^2+k_e^2-1\right)n_{k_e} n_{l_e} -\frac{1}{3}\sum_{k_o < l_o} k_o \left(3 l_o^2+k_o^2-1\right)n_{k_o} n_{l_o}  \nn \\
    & - \frac{1}{3}\sum_{k_e < k_o} k_e \left(3 k_o^2+k_e^2-1\right)n_{k_e} n_{k_o} -\frac{1}{3}\sum_{k_o < k_e} k_o \left(3 k_e^2+k_o^2-1\right)n_{k_o} n_{k_e} \ .
 \label{eq:totalanomSP}
\end{align}

\paragraph{$a$-maximization and the IR SCFT}
  With the anomaly coefficients derived above, we can obtain the IR R-symmetry by maximizing \cite{Intriligator:2003jj} the trial central charge $a(\e)$
  computed from 
    \bea
    R_{{\rm IR}} (\e)
     =     \frac{1 + \e}{2} J_+  + \frac{1 - \e}{2} J_-.
    \eea
  It is important to check that the scalar chiral operators have $R$-charge greater than (or equal to) $2/3$, otherwise unitarity is violated.
  Therefore, we need to know, as the input data, the operator spectrum of $\CT_{{\rm UV}}$, in addition to the central charges $a_{\CT}$, $c_{\CT}$ and $k_F$.
  If we find an operator violating the unitarity bound, we interpret it as being free field along the RG flow and decoupled.
  Thus at the level of the computation of the central charge we subtract its contribution from the trial central charge and redo $a$-maximization,
  as was explained in \cite{Kutasov:2003iy}.
  In our examples we will often see this phenomenon, and the IR theory will be the product of a non-trivial SCFT and a decoupled sector of free fields.

\section{Deformation of conformal SQCD with gauge group $G$}
\label{sec:SQCD} 
  We now apply the generic procedure described in the previous section to the case when $\CT_{{rm UV}}$ is an $\CN=2$ conformal SQCD with gauge group $SU$, $Sp$ and $SO$.
  A busy reader can skip to tables \ref{tab:SU(2)}, \ref{tab:RationalChargesSU} and \ref{tab:RationalChargesSp},
  where the results of the embeddings leading to an IR SCFT with rational central charges are listed.
  All the other embeddings give $\CN=1$ theories with irrational central charges.

\subsection{$G = SU(2) \simeq Sp(1)$, $F=SO(8)$}
\label{subsec:SU(2)}
  We start the study of the $\CN=1$ deformation from the case where $\CT_{{\rm UV}}$ is the $D_4$ theory, 
  namely the $\CN=2$ SCFT realized by $SU(2)$ theory with four fundamental hypermultiplets
  whose flavor symmetry is $SO(8)$. 
  We couple the chiral multiplet $M$ with the $SO(8)$ moment map operator $\mu$ via the superpotential coupling \eqref{superpotential}. 
  We then give $M$ a nilpotent vev corresponding to the embedding $\mathfrak{su}(2) \rightarrow \mathfrak{so}(8)$. 
  
  Let us first review the classification of the embedding. 
  They are classified by a partition of $8$, where even entries appear even number of times. 
  In addition, when the partition is very even, that is, all entries are even numbers, there are two nilpotent orbits. 
  In table \ref{tab:SU(2)}, we tabulate some relevant data for our analysis.
\begin{table}
\centering
\begin{tabular}{|c||c|c|c|c|}
\hline
	Partition & $G_F$ & adjoint & embedding index & IR $\CN=2$ \\
	\hline \hline
	$[7, 1]$ & $\varnothing$ & $V_1 \oplus 2 V_3 \oplus V_5 $ & 28 & Yes; $H_0$ theory \\
	$[5, 3]$ & $\varnothing$ & $3 V_1 \oplus V_2 \oplus 2V_3 $ & 12 & ?  \\
	$[5, 1^3]$ & $\mathfrak{su}(2)$ & $3V_0 \oplus V_1 \oplus 3 V_2 \oplus V_3 $ & 10 & Yes; $H_1$ theory \\
	$[4, 4]^{I, II}$ & $\mathfrak{su}(2)$ &  $3 V_0 \oplus V_1 \oplus 3 V_2 \oplus V_3$ & 10 & Yes; $H_1$ theory \\
	$[3^2, 1^2]$ & $\mathfrak{u}(1)^2$ & $2V_0 \oplus 7 V_1 \oplus V_2$ & 4 & Yes; $H_2$ theory \\
	$[3, 2^2, 1]$ & $\mathfrak{su}(2)$ & $3V_0 \oplus 4 V_{\half} \oplus 3 V_1 \oplus 2 V_{\frac{3}{2}}$ & 3 & No \\
	$[2^4]^{I, II}$ & $\mathfrak{sp}(2)$ & $10 V_0 \oplus 6 V_1 $ & 2 & No \\
	$[3, 1^5]$ & $\mathfrak{sp}(2)$ & $10 V_0 \oplus 6 V_1$ & 2 & No \\
	$[2^2, 1^4]$ & $\mathfrak{su}(2)^3$ & $9 V_0 \oplus 8 V_{\half} \oplus V_1$ & 1 & No \\
	$[1^8] $ & $\mathfrak{so}(8)$ & $28 V_0$ & 0 & Yes; $D_4$ theory \\
	\hline
\end{tabular}
\caption{Classification of the nilpotent vev of $\mathfrak{so}(8)$. 
We also list commuting subalgebra $G_F$ under the embedding of $\mathfrak{su}(2)$ and the decomposition of the adjoint representation, 
along with the embedding index. 
The last Column denotes the supersymmetry enhancement in the IR.
The case with the partition $[1^8]$ simply means no vev, thus the IR theory is the original $\CN=2$ theory plus the free decoupled sector.} 
\label{tab:SU(2)}
\end{table}
   

Let us discuss the RG flow given by each nilpotent vev. 
  
\paragraph{Principal embedding [7, 1]}
  The first line in the table corresponds to the principal embedding where the $SO(8)$ flavor group is broken completely.
  This case was already studied in \cite{Maruyoshi:2016tqk,Maruyoshi:2016aim}, and the IR theory was found to be the ``minimal" (and nontrivial) $\CN=2$ SCFT
  discovered by Argyres and Douglas \cite{Argyres:1995jj}.
  We here briefly review the RG flow in this case, then move on to the other embeddings.
  
  As in table \ref{tab:SU(2)}, the adjoint representation of $SO(8)$ decomposes, under the principal embedding, as
\be
 28 \to V_1 \oplus V_3 \oplus V_5 \oplus V_3 \ .
\ee
  Upon giving the vev to $M$, we are left with $M_{j, -j}$ with $j=1, 3, 5, 3$ with charges $(J_+, J_-) = (0, 4), (0, 8), (0, 12), (0, 8)$. 
The anomaly coefficients after the deformation are given by $\tr J_+ = \tr J_+^3 = -4$, $\tr J_- = 18$, $\tr J_-^3 = 1362$, $\tr J_+^2 J_- = 34$
and $\tr J_+ J_-^2 = - 228$,
from which we get the trial $a$-function as
$a(\e) = - \frac{3}{32}  \left(807 \epsilon ^3-1746 \epsilon ^2+1231 \epsilon -284\right)$. 
Upon $a$-maximization, we get $\e = \frac{1}{807} \left(582+\sqrt{7585}\right) \simeq 0.82911$. 
This makes the Coulomb branch operator (that has $(J_+, J_-)=(0, 4)$) and $M_{j, -j}$ with $j=1$ to violate the unitarity bound so that they become free along the RG flow and get decoupled. 

We redo $a$-maximization after removing these chiral multiplets, and check the dimensions of the remaining chiral operators. This process has to be repeated 
until no operator hits the unitarity bound. 
The final result is that, $\tr \phi^2$, $M_1$, $M_3$ and $M_3'$ decouple.
After removing these operators, the anomaly coefficients are $\tr J_+ = \tr J_+^3 = 0$, $\tr J_- = -2$, $\tr J_-^3 = 622$, $\tr J_+^2 J_- = 14$
and $\tr J_+ J_-^2 = - 112$. This implies $a(\e)= - \frac{3}{32} \left(375 \epsilon ^3-810 \epsilon ^2+559 \epsilon -124\right)$ and
$\e$ is determined to be
\be
 \e = \frac{13}{15} \ , 
\ee
which gives the central charges
\be
 a = \frac{43}{120}, ~~~ 
 c = \frac{11}{30} \ . 
\ee
These are the values of the central charges of the Argyres-Douglas theory \cite{Aharony:2007dj}. 
We also find that the operator $M_{j, -j}$ with $j=5$ has the scalling dimension $\Delta = \frac{6}{5}$,
which is the same as the dimension of the Coulomb branch operator of the Argyres-Douglas theory. 
Therefore we have found an RG flow that takes the $D_4$ theory to the Argyres-Douglas theory (with some free chiral multiplets). 


\paragraph{[5, 3]}
After Higgsing, we have three $M_{1,-1}$, one $M_{2,-2}$ and two $M_{3,-3}$ operators among the components of $M$ whose charges are $(J_+,J_-)=(0,4)$, $(0,6)$ and $(0,8)$. There is no flavor symmetry remaining. 
We find that $M_{1,-1}$ and $M_{2,-2}$ operators (including the Coulomb branch operator having $(J_+,J_-)=(0, 4)$) decouple along the flow. 
At the end of the flow, we find $\e =\frac{41}{51}$ and the two $M_{3,-3}$ operators have the scaling dimension $\Delta = \frac{20}{17}$ and the central charges are
\be
 a = \frac{6349}{13872} \ , \quad c = \frac{3523}{13872} \ . 
\ee
Although the central charges are rational numbers, it does not necessarily mean that the IR interacting theory is $\CN=2$ supersymmetric. 
We are not sure whether this theory is $\CN=1$ or $\CN=2$.

\paragraph{$\mathbf{[5, 1^3] = [4, 4]^I = [4, 4]^{II}}$}
We find that these different choices of nilpotent vevs give rise to the same IR theory. (They have the same decomposition of the adjoint and the same embedding index.) 
 The flavor symmetry is $SU(2)$. After Higgsing, the remaining components of $M$ are three $M_{0,0}$, one $M_{1,-1}$, three $M_{2,-2}$ and one $M_{3,-3}$ fields. Along the flow, $M_{0,0}$, $M_{1,-1}$ and $M_{2,-2}$ operators (and the $\tr \phi^2$) decouple. At the end of the flow, we obtain $\e = \frac{7}{9}$ and the $M_{3,-3}$ operator has the scaling dimension $\Delta = \frac{4}{3}$.
  The central charges are
\be
 a = \frac{11}{24} \ , \quad c = \frac{1}{2} \ . 
\ee
These are exactly the same values as those of the $H_1$ (=$(A_1, A_3)$) theory \cite{Shapere:2008un}, 
which is the maximal conformal point of the $\CN=2$ $SU(2)$ theory with two flavors \cite{Argyres:1995xn}. 
This flow realizes the $\CN=1$ Lagrangian description of the $H_1$ theory.

\paragraph{$\mathbf{[3^2, 1^2]}$}
After Higgsing, we have two $M_{0,0}$, seven $M_{1,-1}$ and one $M_{2,-2}$ components of $M$ remaining. At this point, we have $U(1)^2$ flavor symmetry (in addition to the $U(1)_\CF$). Along the flow,$M_{0,0}$, $M_{1,-1}$ and $\tr \phi^2$ get decoupled. At the end of the flow, we get $\e = \frac{2}{3}$ and the dimension of the $M_{2,-2}$ operator to be $\Delta = \frac{3}{2}$. The central charges are
\be
 a = \frac{7}{12} \ , \quad c = \frac{2}{3} \ , 
\ee
which are precisely the same values as those of the $H_2$ $(=(A_1, D_4))$ theory,
which is the maximal conformal point of the $\CN=2$ $SU(2)$ theory with three flavors \cite{Argyres:1995xn}. 
Thus we propose that the flavor symmetry is also enhanced to $SU(3)$ in the IR.

\paragraph{Other embeddings}
  All the other embeddings give the irrational central charges in the IR. 
  Therefore the theory is $\CN=1$.

\subsection{$G = SU(N)$, $F=SU(2N)$}
\label{sec:LagA1D2N}
  The $\CN=2$ $SU(2)$ conformal SQCD seen in the previous subsection has two generalizations: $SU(N)$ SQCD with $2N$ flavors
  and $Sp(N)$ SQCD with $4N+4$ flavors.
  In this subsection, we consider the former case, namely $\CT =$ $SU(N)$ SQCD.
  To this theory we add a gauge-singlet chiral multiplet $M$. 
  The superpotential of the theory is given by 
\be
W=\Tr \phi \mu + \Tr M \tilde{\mu} \ ,
\label{superpot}
\ee
where $\mu$ and $\tilde{\mu}$ are the moment map operators in the current multiplets for the $SU(N)$ gauge and $SU(2N)$ flavor symmetry respectively. For Lagrangian theories, such as those described here,
they are given by $\mu = Q \widetilde{Q} -\frac{1}{N} \Tr Q \widetilde{Q}$ and $\tilde{\mu} =  \widetilde{Q} Q -\frac{1}{2N} \Tr \widetilde{Q} Q$. 
These theories enjoy an $SU(2N) \times U(1)_B$ flavor symmetry. The chiral superfield $M$ transforms in the adjoint representation of the $SU(2N)$ flavor symmetry.
Upon giving $M$ a nilpotent vev corresponding to various partitions of $2N$, we get a reduced theory whose matter content is the one described in \cite{Agarwal:2014rua}. 

  In \cite{Maruyoshi:2016aim}, the vev corresponding to the so-called principal embedding has been studied. 
  This embedding breaks the $SU(2N)$ symmetry completely and is described by the trivial partition $[2N]$.
  The resulting theory in the IR is the $(A_1, A_{2N-1})$ theory \cite{Cecotti:2010fi}.
  
  In the following we consider the other embeddings, mostly the one corresponding to the partition $[2N-1,1]$
  which breaks the flavor symmetry to $U(1)_m$ and we called as the next-to-maximal embedding. 
  For this embedding, the adjoint representation of $SU(2N)$ decomposes into the representations of $U(1)_m$ 
  (and $SU(2)$)
\be
\begin{split}
	\rm{adj} &\rightarrow \bigoplus_{j=0}^{2N-2}V_{j,0} \ \oplus \ V_{N-1,2N} \ \oplus V_{N-1,-2N} \ .
\end{split}
\ee 
Here $V_{j,q}$ denotes a multiplet that transforms in the spin-$j$ representation of $SU(2)$ and has charge $q$ with respect to $U(1)_m$. 
Upon integrating out massive quarks and removing any decoupled fields, the interacting theory consists of an $SU(N)$ gauge theory with matter fields as given in table \ref{tab:SimpPunctFan}, where $U(1)_1$ and $U(1)_2$ are identified with $U(1)_B$ and $U(1)_m$ respectively.
    \begin{table}
    	\centering
    	\begin{tabular}{|c|c|c|c|c|c|}
    		\hline
    		 fields & $SU(N)$ & $U(1)_1$ & $U(1)_2$ & $J_+$ & $J_-$\\
    		\hline \hline
    		$q_1$ & $\mathbf{N}$ & 1 & $2N-1$ & 1& 0 \\ \hline
    		$\widetilde{q}_1$ & $\bar{\mathbf{N}}$ & $-1$ & $-(2N-1)$ & 1 & 0 \\ \hline
    		$q_2$ & $\mathbf{N}$ & 1 & $-1$ & 1 & $-2(N-1)$ \\ \hline
    		$\widetilde{q}_2$ & $\bar{\mathbf{N}}$  & $-1$ & 1 & 1 & $-2(N-1)$ \\ \hline
    		$\phi$ & \rm{adj} & 0 & 0 & 0 & 2 \\ \hline
    		$M_j$ ~\tiny{$(j=0 , \hdots, 2N-2)$} & $\mathbf{1}$ & 0 & 0 & 0 & $2j+2$ \\ \hline
    		$\widetilde{M}_{N-1}$ & $\mathbf{1}$  & 0 & $2N$ & 0 & $2N$ \\ \hline
    		$\widetilde{M}_{N-1}' $ & $\mathbf{1}$ & 0 & $-2N$ & 0 & $2N$ \\ \hline
    	\end{tabular}
    	\caption{Matter content for the ``Lagrangian description'' of the $(A_1, D_{2N})$ theory.} 
    	\label{tab:SimpPunctFan}
    \end{table}	
The superpotential now becomes 
\be
\begin{split}
  W &= \Tr\widetilde{q}_1 \phi q_1  +   \Tr \widetilde{M}'_{N-1}  q_1 \widetilde{q}_2  +  \Tr \widetilde{M}_{N-1} \widetilde{q}_1 q_2  +  \Tr q_1 \widetilde{q}_1 M_{0} \\
         &\qquad +  \Tr q_2 \widetilde{q}_2 \phi ^{2N-1} + \sum_{j=0}^{2N-2} \sum_{l=0}^{2N-2-j} \Tr q_2 (M_{0})^{l} M_j \widetilde{q}_2 \phi ^{2N-2-j-l} \ .
\end{split}  
\label{eq:SupPotSimpPunct}
\ee
The anomalies of the resulting theory can be easily calculated by using table \ref{tab:SimpPunctFan} and are given by 
\begin{align}
	\begin{split}
		\tr J_+ 
		&=   \tr J_+^3  = -(2N+1), \\
		\tr J_-  &= 2N^2-3,  \\
		\tr J_-^3   
		&= 16 N^4-24 N^3+10 N^2-3, \\
		\tr J_+^2 J_-  &= 6 N^2-3,   \\
		\tr J_+ J_-^2  &= -\frac{32 N^3}{3}+8 N^2+\frac{2 N}{3}-1.
	\end{split}
\end{align}
The central charges $a(\epsilon)$ and $c(\epsilon)$ are therefore given by
\be
\begin{split}
a(\epsilon) = &\left(-\frac{9 N^4}{16}-\frac{9 N^3}{32}-\frac{9 N^2}{64}+\frac{9}{32}\right) \epsilon ^3+\left(\frac{27 N^4}{16}-\frac{45 N^3}{32}-\frac{27 N^2}{64}-\frac{9 N}{32}\right) \epsilon ^2 \\ 
                     &+\left(-\frac{27 N^4}{16}+\frac{117 N^3}{32}-\frac{75 N^2}{64}-\frac{3 N}{16}-\frac{3}{32}\right) \epsilon +\frac{9 N^4}{16}-\frac{63 N^3}{32}+\frac{111 N^2}{64} + \frac{3 N}{32}-\frac{3}{8} \ ,\\
c(\epsilon) = & \left(-\frac{9 N^4}{16}-\frac{9 N^3}{32}-\frac{9 N^2}{64}+\frac{9}{32}\right) \epsilon ^3+\left(\frac{27 N^4}{16}-\frac{45 N^3}{32}-\frac{27 N^2}{64}-\frac{9 N}{32}\right) \epsilon ^2 \\
                     &+\left(-\frac{27 N^4}{16}+\frac{117 N^3}{32}-\frac{71 N^2}{64}-\frac{N}{8}-\frac{5}{32}\right) \epsilon + \frac{9 N^4}{16}-\frac{63 N^3}{32}+\frac{107 N^2}{64} +\frac{5 N}{32}-\frac{1}{4}  \ .             
\end{split}
\ee
Upon $a$-maximization, we find that $\epsilon$ is given by 
\be
\epsilon = \frac{12 N^4-10 N^3-3 N^2-2 N+2 \sqrt{12 N^6+6 N^5+20 N^4-38 N^3+13 N^2+2 N+1}}{3 \left(4 N^4+2 N^3+N^2-2\right)} .\qquad
\label{eq:epsilon1}
\ee 
Using  \eqref{eq:epsilon1} to compute the exact R-charge and hence the dimensions of the various operators in chiral ring of the above theory, we find that the dimensions of the operators $\Tr \phi^k$ with $2 \leq k \leq \lfloor \frac{3N}{4}  \rfloor$
and $M_j$  with $0 \leq j \leq \lfloor \frac{3N}{4}-1 \rfloor$ are less than 1 and decouple. 
After removing these contributions, it might happen that a new set of chiral operators hit the unitarity bound and decouple. We therefore repeat the above cycle until there are no unitarity violating operators in the chiral ring. By explicitly analyzing the above theories for some low values of $N$ (such as $N=2,\hdots, 5$), we find that the end result is that the operators given by $\Tr\phi^k, \ 2 \leq k \leq N$, and $M_j, \ 0 \leq j \leq N-1$, along with $\widetilde{M}_{N-1}$ and $\widetilde{M}'_{N-1}$ decouple from the theory.  After removing these operators from the theory, the corrected $a$-function of the interacting theory therefore becomes 
\be
\begin{split}
  a(\epsilon) &= \left(\frac{9 N^3}{16}-\frac{27 N^4}{64}\right) \epsilon ^3+\left(\frac{81 N^4}{64}-\frac{27 N^3}{8}+\frac{27 N^2}{16}\right) \epsilon ^2 \\
                      &~~+\left(-\frac{81 N^4}{64}+\frac{81 N^3}{16}-\frac{81 N^2}{16}+\frac{3 N}{2}\right) \epsilon + \frac{27 N^4}{64}-\frac{9 N^3}{4}+\frac{27 N^2}{8}-\frac{3 N}{2}.
\end{split}
\label{eq:afinalSU(N)}
\ee
Similarly, the corrected $c$-function is given by 
\be
\begin{split}
c(\epsilon) &= \left(\frac{9 N^3}{16}-\frac{27 N^4}{64}\right) \epsilon ^3+\left(\frac{81 N^4}{64}-\frac{27 N^3}{8}+\frac{27 N^2}{16}\right) \epsilon ^2 \\ 
                     &~~+\left(-\frac{81 N^4}{64}+\frac{81 N^3}{16}-\frac{81 N^2}{16}+\frac{11 N}{8}\right) \epsilon + \frac{27 N^4}{64}-\frac{9 N^3}{4}+\frac{27 N^2}{8}  -\frac{11 N}{8}.
\end{split}
\label{eq:cfinalSU(N)}
\ee
Upon maximizing $a(\epsilon)$ of \eqref{eq:afinalSU(N)}, we find
\be
\epsilon = \frac{3 N-2}{3 N} \ .
\label{eq:finalepsilon}
\ee
By substituting this in \eqref{eq:afinalSU(N)} and \eqref{eq:cfinalSU(N)}, we find that the central charges of the interacting theory are given by 
\be
\begin{split}
a = \frac{6N-5}{12}, ~~~
c = \frac{3N-2}{6} .
\end{split}
\ee
These are exactly what we expect from the central charges of the Argyres-Douglas theories of type $(A_1, D_{2N})$. 
This suggests that the Lagrangian described here flows to the $(A_1,D_{2N})$ theory in the IR. 

Recall that in the set-ups considered in \cite{Maruyoshi:2016aim, Maruyoshi:2016tqk}, the Coulomb branch operators at the IR fixed point arose from the gauge singlet fields that remained coupled to the interacting theory. We find that this is the case for the present theories also. One simple way to see this is to notice that at the fixed point, the dimensions of the fields $M_j , N\leq j \leq 2N-2$, are given by 
\be
\begin{split}
\Delta(M_j) 
            &=\frac{j+1}{N} .  
\end{split}
\ee 
The dimensions of the Coulomb branch operators of the $(A_1,D_{2N})$ theory are given by
\be
\Delta(\CO_i) = 2- \frac{i}{N}, \ i= 1,2, \hdots, N-1\ . 
\ee 
We thus see that the Coulomb branch operator $\CO_i$ is identified with $M_{j=2N-1-i}$. 
 
The $(A_1,D_{2N})$ theory has the $SU(2) \times U(1)$ flavor symmetry with the flavor central charge for the $SU(2)$ symmetry being given by 
\be
k_{SU(2)} = \frac{4(2N-1)}{2N} \ ,
\label{eq:kA1D2N}
\ee
where we have chosen the normalization to be such that a free chiral multiplet transforming as the fundamental of an $SU(N)$ flavor symmetry, contributes 1 to $k_{SU(N)}$. It must be that in the IR of theory described in this section, a linear combination of the two $U(1)$ flavor symmetries, gets enhanced to the $SU(2) \subset SU(2) \times U(1)$ flavor symmetry present in the $(A_1,D_{2N})$ theory. This linear combination can be easily obtained by requiring that the corresponding central charge matches with the value given in \eqref{eq:kA1D2N}. In order to do this, we remind ourselves of the following facts: For systems with $\CN=2$ supersymmetry and a flavor symmetry $G$, the central charge $k_G$ is related to the 't Hooft anomaly via the relation
\be
k_G \delta^{ab} = -2\Tr R_{\CN=2} T^a T^b.
\ee
The $\CN=1$ R-symmetry inside $\CN=2$ superconformal algebra is given by 
\be
R_{\CN=1} = \frac{1}{3} R_{\CN=2} + \frac{4}{3} I_3.
\ee 
It therefore follows that 
\be
k_G \delta^{ab} = - 6\Tr R_{\CN=1} T^a T^b \ ,
\label{eq:FlavorCentralCharge}
\ee

Now, let the linear combination we are interested in be $\a_1 U(1)_1 + \a_2 U(1)_2 $, where $\a_1, \a_2$ are constants that we have to determine. The corresponding central charge $k(\a_1,\a_2)$ is then given by
\be
\begin{split}
k(\a_1,\a_2) 
           &=-12 N (-2 \a_1^2 N + 2 \a_2^2 N -4 \a_2^2 N^2)\Big(\frac{1-\epsilon}{2}\Big) \ .
\end{split}
\label{eq:CentralChargeA1D2N}
\ee  
In the above, we did not include the contribution of $\widetilde{M}, \ \widetilde{M}'$ in $k(\a_1,\a_2)$ in accordance with the fact that, somewhere along the RG flow, they decouple from the interacting theory. Now by substituting from \eqref{eq:finalepsilon} in \eqref{eq:CentralChargeA1D2N}, we get
\be
k(\a_1,\a_2) = 8N (2N \a_2^2 - \a_2^2 + \a_1^2) \ .
\ee
Comparing this with the central charge given in \eqref{eq:kA1D2N}, we see that there is a one-parameter space of solutions for $\a_1$ and $\a_2$. Of all these, a particularly simple solution is given by 
\be
\begin{split} 
\a_1&=0 \ , ~~~ \a_2 = \frac{1}{2N} \ .
\label{eq:A1D2Nflv} 
\end{split} 
\ee 

\paragraph{Other possible nilpotent vevs for $M$} 
So far we have only focussed on the next-to-maximal embedding. The corresponding RG flow was shown to bring the theory to the fixed point describing the $(A_1, D_{2N})$ theory. 
At this point, it is quite natural to hope that the other nilpotent embeddings of $SU(2N)$, also trigger RG flows to various other AD theories. Unfortunately, this does not seem to be the case. We scanned through the set of nilpotent embedding of $SU(2N)$ for explicit values of $N$, $2 \leq N \leq 10$, and found that in most of the cases the central charges of the IR theory are irrational numbers. It was argued in \cite{Maruyoshi:2016aim} that the central charges of 4d $\CN=2$ theories must be rational. It therefore follows the nilpotent embeddings with irrational IR central charges do not experience any supersymmetry enhancement. There are sporadic cases where the IR central charges do end up being rational, however, neither were we able to find a definite pattern underlying such embeddings nor did the corresponding central charges seem to fit those of the various generalized AD theories in literature so far. We are therefore not sure if these cases genuinely lead to an IR theory with the enhanced supersymmetry. For the sake of completeness, we list the nilpotent embeddings leading to rational IR central charges in table \ref{tab:RationalChargesSU}.

\newpage
\begin{center}    
	\begin{longtable}{|c|c|c|c|c|c|}
		\caption{Nilpotent embeddings of $SU(2N)$ leading to rational values for $a$ and $c$. The partition $[1^{2N}]$ reduces to four-dimensional $\CN=2$ $SU(N)$ gauge theory with $2N$ hypermultiplets plus decoupled fields. The partitions $[2N-1,1]$ and $[2N]$ reduce to AD theories listed in the last column of the above table.} \\ 
		\hline
		$SU(2N)$ & $\rho:SU(2) \hookrightarrow SU(2N)$ & $a$ & $c$ & 4d $\CN=2$ SUSY \\ 
		\hline \hline
		\multirow{3}{*}{$SU(4)$} & $[1^{4}]$ & $\frac{23}{24}$ & $\frac{7}{6}$ & Yes; $N_c=2, \ N_f=4$ \\
		\cline{2-5}	
		& $[3,1]$ & $\frac{7}{12}$ & $\frac{2}{3}$ &Yes; $(A_1,D_{4})$ AD th. \\
		\cline{2-5}	
		& $[4]$ & $\frac{11}{24}$ & $\frac{1}{2}$ &Yes; $(A_1,A_{3})$ AD th.  \\
		\hline \hline
		\multirow{3}{*}{$SU(6)$} & $[1^{6}]$ & $\frac{29}{12}$ & $\frac{17}{6}$ & Yes; $N_c=3, \ N_f=6$ \\
		\cline{2-5}	
		& $[5,1]$ & $\frac{13}{12}$ & $\frac{7}{6}$ &Yes; $(A_1,D_{6})$ AD th. \\
		\cline{2-5}	
		& $[6]$ & $\frac{11}{12}$ & $\frac{23}{24}$ &Yes; $(A_1,A_{5})$ AD th. \\
		\hline \hline
		\multirow{5}{*}{$SU(8)$} & $[1^{8}]$ & $\frac{107}{24}$ & $\frac{31}{6}$ & Yes; $N_c=4, \ N_f=8$ \\
		\cline{2-5}	
		& $[2,1^6]$ & $\frac{73801}{17424}$ & $\frac{43121}{8712}$ &? \\
		\cline{2-5}	
		& $[4,4]$ & $\frac{9097}{3888}$ & $\frac{5129}{1944}$ & ? \\
		\cline{2-5}	
		& $[7,1]$ & $\frac{19}{12}$ & $\frac{5}{3}$ &Yes; $(A_1,D_{8})$ AD th. \\
		\cline{2-5}	
		& $[8]$ & $\frac{167}{120}$ & $\frac{43}{30}$ &Yes; $(A_1,A_{7})$ AD th. \\
		\hline\hline
		\multirow{5}{*}{$SU(10)$} & $[1^{10}]$ & $\frac{247}{24}$ & $\frac{71}{6}$ & Yes; $N_c=5, \ N_f=10$ \\
		\cline{2-5}	
		& $[5,1^5]$ & $\frac{5553943}{1383123}$ & $\frac{6257387}{1383123}$ &$?$ \\
		\cline{2-5}	
		& $[5,3,1^2]$ & $\frac{92540867}{24401712}$ & $\frac{52091009}{12200856}$ &$?$ \\
		\cline{2-5}	
		& $[9,1]$ & $\frac{25}{12}$ & $\frac{13}{6}$ &Yes; $(A_1,D_{10})$ AD th. \\
		\cline{2-5}	
		& $[10]$ & $\frac{15}{8}$ & $\frac{23}{12}$ &Yes; $(A_1,A_{9})$ AD th. \\
		\hline\hline
		\multirow{4}{*}{$SU(12)$} & $[1^{12}]$ & $\frac{247}{24}$ & $\frac{71}{6}$ & Yes; $N_c=6, \ N_f=12$ \\
		\cline{2-5}	
		& $[4^3]$ & $\frac{754501}{138384}$ & $\frac{424727}{69192}$ &$?$ \\
		\cline{2-5}	
		& $[11,1]$ & $\frac{31}{12}$ & $\frac{8}{3}$ &Yes; $(A_1,D_{12})$ AD th. \\
		\cline{2-5}			
		& $[12]$ & $\frac{397}{168}$ & $\frac{101}{42}$ &Yes; $(A_1,A_{11})$ AD th. \\
		\hline \newpage\hline
		$SU(2N)$ & $\rho:SU(2) \hookrightarrow SU(2N)$ & $a$ & $c$ & 4d $\CN=2$ SUSY \\ 
		\hline \hline
		\multirow{3}{*}{$SU(14)$} & $[1^{14}]$ & $\frac{169}{12}$ & $\frac{97}{6}$ & Yes; $N_c=7, \ N_f=14$ \\
		\cline{2-5}	
		& $[13,1]$ & $\frac{37}{12}$ & $\frac{19}{6}$ &Yes; $(A_1,D_{14})$ AD th. \\
		\cline{2-5}	
		& $[14]$ & $\frac{137}{48}$ & $\frac{139}{48}$ &Yes; $(A_1,A_{13})$ AD th. \\
		\hline \hline	
		\multirow{10}{*}{$SU(16)$} & $[1^{16}]$ & $\frac{443}{24}$ & $\frac{127}{6}$ & Yes; $N_c=8, \ N_f=18$ \\
		\cline{2-5}
		& $[2^2,1^{12}]$ & $\frac{630199}{35912}$ & $\frac{364579}{17956}$ &$?$ \\
		\cline{2-5}
		& $[4,2^4,1^{4}]$ & $\frac{25431}{1936}$ & $\frac{14525}{968}$ &$?$ \\
		\cline{2-5}
		& $[4,2^6]$ & $\frac{228913}{17328}$ & $\frac{134131}{8664}$ &$?$ \\
		\cline{2-5}
		& $[4^4]$ & $\frac{214899}{21904}$ & $\frac{120899}{10952}$ &$?$ \\
		\cline{2-5}
		& $[6,2,1^8]$ & $\frac{116437}{11664}$ & $\frac{64523}{5832}$ &$?$ \\
		\cline{2-5}
		& $[7,3^2,1^3]$ & $\frac{149}{18}$ & $\frac{82}{9}$ &$?$ \\
		\cline{2-5}
		& $[7,4^2,1]$ & $\frac{1127683}{142884}$ & $\frac{625739}{71442}$ &$?$ \\
		\cline{2-5}
		& $[12,2^2]$ & $\frac{251183}{51984}$ & $\frac{136621}{25992}$ &$?$ \\
		\cline{2-5}
		& $[15,1]$ & $\frac{43}{12}$ & $\frac{11}{3}$ &Yes; $(A_1,D_{16})$ AD th. \\
		\cline{2-5}
		& $[16]$ & $\frac{241}{72}$ & $\frac{61}{18}$ & Yes; $(A_1,A_{15})$ AD th. \\
		\hline \hline
		\multirow{3}{*}{$SU(18)$}  & $[1^{18}]$ &$\frac{281}{12}$ & $\frac{161}{6}$ & Yes; $N_c=9, \ N_f = 18$ \\
		\cline{2-5}
		& $[17,1]$ & $\frac{49}{12}$ & $\frac{25}{6}$ & Yes; $(A_1,D_{18})$ AD th. \\
		\cline{2-5}
		& $[18]$ & $\frac{461}{120}$ & $\frac{233}{60}$ & Yes; $(A_1,A_{17})$ AD th. \\
		\hline\hline	
		\multirow{8}{*}{$SU(20)$} & $[1^{20}]$& $\frac{627}{24}$ & $\frac{199}{6}$ & Yes; $N_c=10, \  N_f =20$ \\ 
		\cline{2-5}
		& $[4^5]$ & $\frac{207429}{13456}$ & $\frac{116663}{6728}$ & $?$\\
		\cline{2-5}
		& $[5^2, 1^{10}]$ & $\frac{60991789}{3682992}$ & $\frac{34227083}{1841496}$ & $?$\\
		\cline{2-5}
		&$[5^2,3^2,1^4]$ & $\frac{1585050209}{101338032}$ & $\frac{890667263}{50669016}$ & $?$\\
		\cline{2-5}
		&$[6,2^7]$ & $\frac{954706769}{59889072}$ & $\frac{528458927}{29944536}$ & $?$\\
		\cline{2-5}
		&$[7,6,3,2,1^2]$ & $\frac{100877}{8112}$ & $\frac{56243}{4056}$ & $?$\\
		\cline{2-5}
		&$[19,1]$ & $\frac{55}{12}$ & $\frac{14}{3}$ & Yes; $(A_1,D_{20})$ AD th. \\
		\cline{2-5}
		&$[20]$ & $\frac{1145}{264}$ & $\frac{289}{66}$ & Yes; $(A_1,A_{19})$ AD th.  \label{tab:RationalChargesSU} \\
		\hline 															
	\end{longtable}
\end{center}


\subsection{$G = Sp(N)$, $F=SO(4N+4)$}
\label{sec:LagA1D2N+1}
  In this subsection, we will describe Lagrangians that flow to the $(A_1, D_{2N+1})$ theories. 
  These are obtained by considering the $\CN=1$ deformation taking $\CT_{{\rm UV}}$ 
  as $\CN=2$ $Sp(N)$ gauge theory with $4N+4$ fundamental half-hypermultiplets $Q$ whose flavor symmetry is $F=SO(4N+4)$.  
  Prior to giving any expectation value to $M$ in the adjoint representation of $SO(4N+4)$, the superpotential is in the same form as \eqref{superpot}, 
where $\mu = Q Q^t -\frac{1}{N} \Tr Q Q^t$ and $\tilde{\mu} =  Q^t Q -\frac{1}{4N+4} \Tr Q^tQ$, 
with appropriate insertions of the two-index antisymmetric tensor, $\Omega$, that is invariant under transformations by elements of the $Sp(N)$ group. 
We now give a nilpotent vev to $M$. 
These are classified by $\mathfrak{su}(2) \hookrightarrow \mathfrak{so}(4N+4)$ and are labelled by the partitions of $4N+4$ for which even parts occur with even multiplicity \cite{collingwood17008nilpotent}.

In particular, here we are interested in the nilpotent vev corresponding to the partition given by $[4N+1, 1^3]$.
This leaves an $SO(3) \subset SO(4N+4)$  unbroken. Accordingly, the adjoint representation of $SO(4N+4)$ decomposes into irreps. of $SU(2) \times SO(3)$ as
\be
{\rm adj} \rightarrow \bigoplus_{k=0}^{2N-1} (V_{2k+1} \otimes \mathbf{1} ) \oplus (V_{2N} \otimes \mathbf{3})   \oplus (V_0 \otimes \mathbf{3}) \ .
\ee
At energies below the scale of the vev $\vev{M}$, the matter content of the theory is given by the fields enumerated in table \ref{tab:OrthFanAD}.
\begin{table}
    	\centering
    	\begin{tabular}{|c|c|c|c|c|}
    		\hline
    		 fields & $Sp(N)$ & $SO(3)$  & $J_+$ & $J_-$\\
    		\hline \hline
    		$q_1$ & $\mathbf{2N}$ & $\mathbf{3}$  & 1& 0 \\ \hline
    		$q_2$ & $\mathbf{2N}$ & $\mathbf{1}$ &  1 & $-4N$ \\ \hline
    		$\phi$ & \rm{adj} & $\mathbf{1}$  & 0 & 2 \\ \hline
    		$M_{j=2k+1}$,~{ $(0 \leq k \leq 2N-1)$} & $\mathbf{1}$ & $\mathbf{1}$ & 0 & $2j+2$ \\ \hline	
    		$M_0$ & $\mathbf{1}$ & $\mathbf{3}$ & 0 & $2$ \\ \hline
		$M_{2N}$ & $\mathbf{1}$  & $\mathbf{3}$ &  0 & $4N+2 $ \\ \hline
    	\end{tabular}
    	\caption{Matter content of the ``Lagrangian description'' for the $(A_1, D_{2N+1})$ theory.} 
    	\label{tab:OrthFanAD}
\end{table}
The corresponding low energy superpotential (up to appropriate insertions of $\Omega$) becomes
\begin{align}
\label{eq:SupPotOrthFanAD}
\begin{split}
W = \Tr q_1 \phi q_1 + \Tr M_{2N} q_1 q_2 + \Tr M_0 q_1 q_1 + \Tr q_2  \phi^{4N+1} q_2 + \sum_{ k=0}^{2N-1} \Tr q_2 M_j q_2 \phi ^{4N-j}\Big|_{j=2k+1} \ .
\end{split}
\end{align}
The anomalies of the resulting theory are now given by
\begin{align}
	\begin{split}
		\tr J_+ &= \tr J_+^3    
		= -(2N+6), \\
		\tr J_- &= 4N^2 + 8N +6,   \\
		\tr J_-^3 
		&= 128 N^4+224 N^3+116 N^2+24 N+6, \\
		\tr J_+^2 J_-  &= 12 N^2+16 N+6,  \\
		\tr J_+ J_-^2  &= -\frac{128 N^3}{3}-64 N^2-\frac{70 N}{3}-6.
	\end{split}
\end{align}
The central charges $a(\epsilon)$ and $c(\epsilon)$ are found to be
    \bea
    a(\epsilon)
    &=&    \left(-\frac{9 N^4}{2}-\frac{99 N^3}{8}-\frac{387 N^2}{32}-\frac{81 N}{16}-\frac{27}{16}\right) \epsilon ^3+\left(\frac{27 N^4}{2}+\frac{225 N^3}{8}+\frac{567 N^2}{32}+\frac{99 N}{32}\right) \epsilon^2 \nonumber \\ 
    &{ }&-\left(\frac{27 N^4}{2}+\frac{153 N^3}{8}+\frac{129 N^2}{32}-\frac{15 N}{8}-\frac{9}{16}\right) \epsilon + \frac{9 N^4}{2}+\frac{27 N^3}{8}-\frac{51 N^2}{32}-\frac{9 N}{32}, \nonumber \\
    c(\epsilon)
    &=&    \left(-\frac{9 N^4}{2}-\frac{99 N^3}{8}-\frac{387 N^2}{32}-\frac{81 N}{16}-\frac{27}{16}\right) \epsilon ^3+\left(\frac{27 N^4}{2}+\frac{225 N^3}{8}+\frac{567 N^2}{32}+\frac{99 N}{32}\right) \epsilon^2 \nonumber \\
    &{ }&- \left(\frac{27 N^4}{2}+\frac{153 N^3}{8}+\frac{125 N^2}{32}-\frac{35 N}{16}-\frac{15}{16}\right) \epsilon+\frac{9 N^4}{2}+\frac{27 N^3}{8}-\frac{55 N^2}{32} -\frac{15 N}{32} .       	          
    \label{eq:OrthFanADpass1}
    \eea 
We now fix $\epsilon$ to be the value that maximizes $a(\epsilon)$. This is given by
\begin{align}
\begin{split}
\epsilon
&=    \frac{48 N^4+100 N^3+63 N^2+11 N}{48 N^4+132 N^3+129 N^2+54 N+18}  \\
&\qquad   +\frac{\sqrt{192 N^6+720 N^5+760 N^4+512 N^3+481 N^2+228 N+36}}{48 N^4+132 N^3+129 N^2+54 N+18} \ . 
\label{eq:OrthFanEpsilonPass1}
\end{split}
\end{align}
As in the previous case, with this the value of $\epsilon$ we find that some of chiral operators violate the unitarity bound and hence must have decoupled from the interacting theory. 
We therefore remove them from the spectrum and redo a-maximization using the corrected central charges and repeat this cycle until there are no gauge invariant chiral operators that violate the unitarity bound. 
By explicitly checking for low values of $N$, such as $1 \leq N \leq 5$, we conclude that in the end, the chiral operators $\Tr \phi^{2k}$ with $1 \leq k \leq N$ and $M_j$ with $j= 2k+1, \ 0 \leq k \leq N-1 $, along with $ M_0$ and $M_{2N}$, decouple from the interacting theory. The correct central charges of the interacting theory at its IR fixed point are therefore given by
\be
\begin{split}
	a(\epsilon) =& \left(-\frac{27 N^4}{8}-\frac{27 N^3}{8}-\frac{27 N^2}{32}\right) \epsilon ^3+\left(\frac{81 N^4}{8}+\frac{27 N^3}{8}-\frac{81 N^2}{32}-\frac{27 N}{32}\right) \epsilon ^2 \\ 
            	&+ \left(-\frac{81 N^4}{8}+\frac{27 N^3}{8}+\frac{135 N^2}{32}+\frac{9 N}{16}\right) \epsilon + \frac{27 N^4}{8}-\frac{27 N^3}{8}-\frac{27 N^2}{32}+\frac{9 N}{32} \ , \\
	c(\epsilon) =&\left(-\frac{27 N^4}{8}-\frac{27 N^3}{8}-\frac{27 N^2}{32}\right) \epsilon ^3+\left(\frac{81 N^4}{8}+\frac{27 N^3}{8}-\frac{81 N^2}{32}-\frac{27 N}{32}\right) \epsilon ^2\\ 
	&+ \left(-\frac{81 N^4}{8}+\frac{27 N^3}{8}+\frac{135 N^2}{32}+\frac{3 N}{8}\right) \epsilon +\frac{27 N^4}{8}-\frac{27 N^3}{8}-\frac{27 N^2}{32}+\frac{15 N}{32}.
\end{split}
\label{eq:OrthFanADfinal}
\ee 
Upon maximizing the central charge, $a(\epsilon)$, in \eqref{eq:OrthFanADfinal}, we find that $\epsilon$ is given by
\be
\epsilon = \frac{6 N+1}{6 N+3},
\label{eq:OrthFanEpsilonFinal}
\ee  
from which we get the central charges
\bea
	a = \frac{N (8 N+3)}{16 N+8},~~~
	c = \frac{N}{2}.
\label{eq:AD2N+1}
\eea 
These are identical to those of the $(A_1,D_{2N+1})$ theory. 
We therefore conjecture that the interacting theory hereby obtained by us, is identical to the $(A_1,D_{2N+1})$ theory. 
Our conjecture is also supported by the fact that the Coulomb branch operator dimensions of $(A_1,D_{2N+1})$ are given by 
\be
\Delta(\CO_i) = 2 -\frac{2 i}{2N+1}, \ 1 \leq i \leq N \ .
\label{dima1ad2N+1}
\ee 
This spectrum is beautifully reproduced from the interacting theory at the IR fixed point of the theories being considered here:
the dimensions of the operators $M_j$ with $j =2k+1, N \leq k \leq 2N-1 $ are given by 
\be
\begin{split} 
\Delta(M_j) &= \frac{2k+2}{2N+1}, ~~~ j=2k+1, \ N\leq k \leq 2N-1 \ .  
\end{split} 
\ee 
Writing, $k=2N-i, \ 1 \leq i \leq N$, we see the equivalence with \eqref{dima1ad2N+1}.
We therefore find that the dimension of the $i$-th Coulomb branch operator, $\CO_i$, in the $(A_1,D_{2N+1})$ theory agrees with that of $M_{j=4N-2i+1}$ in the theory considered here.

 We also notice that our set-up has an $SO(3)$ global symmetry while the $(A_1, D_{2N+1})$ theories have an $SU(2)$ global symmetry. The equivalence between the Lie algebra of $SO(3)$ and $SU(2)$ therefore lends further evidence to our conjecture. To see that indeed the $SO(3)$ global symmetry of our theories matches with that of the $(A_1, D_{2N+1})$ theories, we now show that their flavor central charges agree. Recall, from \eqref{eq:FlavorCentralCharge}, that the flavor central charge is given by
\be
k_{SO(3)} \delta^{ab} = - 6\Tr R_{\CN=1} T^a T^b \ ,
\ee
where $T^a$, $T^b$ are the generators of the $SO(3)$ global symmetry of our theory. Since $M_0$ and $M_{2N}$, ultimately decouple, therefore the only fields that contribute to $k_{SO(3)}$ are those in the multiplet $q_1$. By substituting $\epsilon$ from \eqref{eq:OrthFanEpsilonFinal} to compute the exact $\CN=1$ superconformal R-charge of the theory, we find 
\be
\begin{split}
k_{SO(3)} 
          &=\frac{8N}{2N+1},
\end{split}          
\ee
which agrees with the flavor central charge $k_{SU(2)}$ of the $(A_1,D_{2N+1})$ theory. 

\paragraph{Other possible nilpotent vevs for $M$} Similar to the case of $SU(N)$ gauge theory in the previous section, there seems to be no other nilpotent embeddings which trigger flows with the supersymmetry enhancement in the IR in the case of $Sp(N)$ gauge theories. Once again, there are sporadic cases where the central charges $a$ and $c$ become rational, however they do not seem to correspond to any known $\CN=2$ SCFTs. For completeness, we list the corresponding partitions of $SO(4N+4)$ flavor symmetry in table \ref{tab:RationalChargesSp}.

\begin{center}    
	\begin{longtable}{|c|c|c|c|c|c|}
		\caption{Nilpotent embeddings of $SO(4N+4)$ leading to rational values for $a$ and $c$. 
		The partition $[1^{4N+4}]$ reduces to four-dimensional $\CN=2$ $Sp(N)$ gauge theory with $4N+4$ half-hypermultiplets plus decouped fields. The partitions $[4N+1,1^3]$ and $[4N+4]$ reduce to AD theories listed in the last column of the above table. Here $N_c$ denotes the rank of the corresponding symplectic gauge group and $N_f$ denotes the number of half-hypermultiplets transforming in the fundamental representation of the gauge group.} \\ 
		\hline
		$SO(4N+4)$ & $\rho:SU(2) \hookrightarrow SO(4N+4)$ & $a$ & $c$ & 4d $\CN=2$ SUSY \\ 
		\hline \hline
		\multirow{5}{*}{$SO(8)$} & $[1^{8}]$ & $\frac{23}{24}$ & $\frac{7}{6}$ & Yes; $N_c=1, \ N_f=8$ \\
		\cline{2-5}	
		& $[3^2,1^2]$ & $\frac{7}{12}$ & $\frac{2}{3}$ &Yes; $(A_1,D_{4})$ AD th.  \\
		\cline{2-5}	
		& $[4,4] \equiv [5,1^3]$ & $\frac{11}{24}$ & $\frac{1}{2}$ & Yes; $(A_1,D_{3})$ AD th. \\
		\cline{2-5}	
		& $[5,3]$ & $\frac{6349}{13872}$ & $\frac{3523}{6936}$ &? \\
		\cline{2-5}	
		& $[7,1]$ & $\frac{43}{120}$ & $\frac{11}{30}$ &Yes; $(A_1,A_2)$ AD th. \\
		\hline\hline
		\multirow{4}{*}{$SO(12)$} & $[1^{12}]$ & $\frac{37}{12}$ & $\frac{11}{3}$ & Yes; $N_c=2, \ N_f=12$ \\
		\cline{2-5}	
		& $[4^2,2^2]$ & $\frac{105027}{59536}$ & $\frac{61145}{29768}$ &? \\
		\cline{2-5}	
		& $[9,1^3]$ & $\frac{19}{20}$ & $1$ &Yes; $(A_1,D_5)$ AD th. \\
		\cline{2-5}	
		& $[11,1]$ & $\frac{67}{84}$ & $\frac{17}{21}$ &Yes; $(A_1,A_4)$ AD th. \\
		\hline\hline
		\multirow{5}{*}{$SO(16)$} & $[1^{16}]$ & $\frac{51}{8}$ & $\frac{15}{2}$ & Yes; $N_c=3, \ N_f=16$ \\
		\cline{2-5}	
		& $[5,1^{11}]$ & $\frac{109031}{27744}$ & $\frac{123889}{27744}$ &? \\
		\cline{2-5}	
		& $[5,3^3,1^2]$ & $\frac{18250741}{5195568}$ & $\frac{10440877}{2597784}$ &? \\
		\cline{2-5}	
		& $[13,1^3]$ & $\frac{81}{56}$ & $\frac{3}{2}$ & Yes; $(A_1,D_{7})$ AD th. \\
		\cline{2-5}	
		& $[15,1]$ & $\frac{91}{72}$ & $\frac{23}{18}$ &Yes; $(A_1,A_6)$ AD th. \\
		\hline\hline
		\multirow{3}{*}{$SO(20)$} & $[1^{20}]$ & $\frac{65}{6}$ & $\frac{38}{3}$ & Yes; $N_c=4, \ N_f=20$ \\
		\cline{2-5}	
		& $[2^2,1^{16}]$ & $\frac{4181}{400}$ & $\frac{2463}{200}$ &? \\
		\cline{2-5}	
		& $[3^4,2^4]$ & $\frac{29}{4}$ & $\frac{133}{16}$ &? \\
		\hline
		$SO(4N+4)$ & $\rho:SU(2) \hookrightarrow SO(4N+4)$ & $a$ & $c$ & 4d $\CN=2$ SUSY \\ 
		\hline \hline
		\multirow{10}{*}{$SO(20)$}& $[4^4,2^2]$ & $\frac{28361329}{4702512}$ & $\frac{16338643}{2351256}$ &? \\
		\cline{2-5}	
		& $[9,5,3,1^3]$ & $\frac{737}{192}$ & $\frac{817}{192}$ &? \\
		\cline{2-5}	
		& $[11,1^9]$ & $\frac{6638927}{1976856}$ & $\frac{3700169}{988428}$ &? \\
		\cline{2-5}	
		& $[11,2^2,1^5]$ & $\frac{106413731}{31795224}$ & $\frac{59339969}{15897612}$ &? \\
		\cline{2-5}	
		& $[11,2^4,1] \equiv [11,3,1^6] $ & $\frac{26650955}{7990296}$ & $\frac{14869241}{3995148}$ &? \\
		\cline{2-5}	
		& $[11,3,2^2,1^2] $ & $\frac{106793099}{32127576}$ & $\frac{59613689}{16063788}$ &? \\
		\cline{2-5}
		& $[11,3^2,1^3] $ & $\frac{1671587}{504600}$ & $\frac{233399}{63075}$ &? \\
		\cline{2-5}
		& $[11,3^3] $ & $\frac{26839019}{8157336}$ & $\frac{15005561}{4078668}$ &? \\
		\cline{2-5}
		& $[17,1^3]$ & $\frac{35}{18}$ & $2$ & Yes; $(A_1,D_{9})$ AD th. \\
		\cline{2-5}	
		& $[19,1]$ & $\frac{115}{66}$ & $\frac{58}{33}$ &Yes; $(A_1,A_8)$ AD th. \\
		\hline\hline
		\multirow{11}{*}{$SO(24)$} & $[1^{24}]$ & $\frac{395}{24}$ & $\frac{115}{6}$ & Yes; $N_c=5, \ N_f=24$ \\
		\cline{2-5}	
		& $[4^2, 2^6,1^4]$ & $\frac{511}{48}$ & $\frac{289}{24}$ &? \\
		\cline{2-5}		
		& $[4^4,3, 2^2,1]$ & $\frac{1092067}{115248}$ & $\frac{630289}{57624}$ &? \\
		\cline{2-5}	
		& $[5, 2^8,1^3]$ & $\frac{83}{8}$ & $\frac{187}{16}$ &? \\
		\cline{2-5}	
		& $[5,3^2,2^4,1^5]$ & $\frac{70571}{6936}$ & $\frac{40037}{3468}$ &? \\
		\cline{2-5}	
		& $[5^2, 2^4,1^6]$ & $\frac{19241078}{2031987}$ & $\frac{21995659}{2031987}$ &? \\
		\cline{2-5}	
		& $[5^2, 3,1^{11}]$ & $\frac{5461835}{578888}$ & $\frac{3107921}{289444}$ &? \\
		\cline{2-5}	
		& $[5^2, 3^3,1^{8}]$ & $\frac{18737545}{2036928}$ & $\frac{21328613}{20136928}$ &? \\
		\cline{2-5}	
		& $[11,7,5,1]$ & $\frac{5753}{1200}$ & $\frac{3179}{600}$ &? \\
		\cline{2-5}	
		& $[21,1^3]$ & $\frac{215}{88}$ & $\frac{5}{2}$ & Yes; $(A_1,D_{11})$ AD th. \\
		\cline{2-5}
		& $[23,1]$ & $\frac{695}{312}$ & $\frac{175}{78}$ &  Yes; $(A_1,A_{10})$ AD th. \label{tab:RationalChargesSp} \\
		\hline														
	\end{longtable}
\end{center}

\subsection{$G=SO(N)$, $F=Sp(N-2)$}
\label{sec:SO(N)deform}

Finally we consider the IR theories obtained by deforming $\CN=2$ $SO(N)$ gauge theory coupled to $2N-4$ fundamental half-hypermultiplets. The flavor symmetry in this case is given by $Sp(N-2)$ and the deformation is given by coupling a gauge singlet field $M$ in the adjoint representation of $Sp(N-2)$, to the moment map operator of $Sp(N-2)$, and then giving $M$ a nilpotent vev. By Jacobson and Morozov's theorem, these are in one-to-one correspondence with those partitions of $2N-4$ for which the odd parts occur with even multiplicity. The anomalies of the deformed theory can be obtained from \eqref{eq:totalanomSP}. We find that for most of the nilpotent vevs of $M$, the central charges of the IR theories are irrational. Occasionally, it happens that there is a nilpotent vev for which the central charges become rational. These are listed in table \ref{tab:RationalChargesSO}.  There appears to be no definite organizing principle behind the cases for which the nilpotent vev leads to rational central charges. Neither were we able to find any $\CN=2$ theories whose central charges would agree with those listed in table \ref{tab:RationalChargesSO}.

\begin{center}    
	\begin{longtable}{|c|c|c|c|c|c|}
		\caption{Nilpotent embeddings of $Sp(N_c-2)$ (s.t. $Sp(1)\simeq SU(2)$) leading to rational values for $a$ and $c$. 
		The partition $[1^{2N_c-4}]$ reduces to four-dimensional $\CN=2$ $SO(N_c)$ gauge theory with $N_f=2N_c-4$ half-hypermultiplets. } \\ 		\hline
		$Sp(N-2)$ & $\rho:SU(2) \hookrightarrow Sp(N-2)$ & $a$ & $c$ & 4d $\CN=2$ SUSY \\ 
		\hline \hline
		\multirow{2}{*}{$Sp(2)$} & $[1^{4}]$ & $\frac{19}{12}$ & $\frac{5}{3}$ & Yes; $N_c=4, \ N_f=4$ \\
		\cline{2-5}	
		& $[2,1^2]$ & $\frac{10111}{7056}$ & $\frac{5381}{3528}$ &? \\
		\hline \hline	
		\multirow{2}{*}{$Sp(3)$} & $[1^{6}]$ & $\frac{65}{24}$ & $\frac{35}{12}$ & Yes; $N_c=5, \ N_f=6$ \\
		\cline{2-5}
		 & $[4,1^2]$ & $\frac{325}{192}$ & $\frac{341}{192}$ &? \\
		\hline \hline
		\multirow{1}{*}{$Sp(4)$} & $[1^{8}]$ & $\frac{33}{8}$ & $\frac{9}{2}$ & Yes; $N_c=6, \ N_f=8$ \\
                 \hline \hline		
		\multirow{1}{*}{$Sp(5)$} & $[1^{10}]$ & $\frac{35}{6}$ & $\frac{77}{12}$ & Yes; $N_c=7, \ N_f=10$ \\
		\hline \hline
		\multirow{3}{*}{$Sp(6)$} & $[1^{12}] $ & $\frac{47}{6}$ & $\frac{26}{3}$ & Yes; $N_c=8, \ N_f=12$\\
		\cline{2-5}
		& $[2^2,1^{8}]$ & $\frac{589093}{80688}$ & $\frac{329335}{40344}$ &? \\
		\cline{2-5}
		& $[4,1^8]$ & $\frac{13065}{2312}$ & $\frac{7085}{1156}$ & ? \\
		\hline \hline
		\multirow{3}{*}{$Sp(7)$} & $[1^{14}]$ & $\frac{81}{8}$ & $\frac{45}{4}$ & Yes; $N_c=9, \ N_f=14$ \\
		\cline{2-5}	
		& $[5^2,1^4]$ & $\frac{59094550}{10978707}$ & $\frac{129141025}{21957414}$ &? \\
		\cline{2-5}		
		& $[6, 3^2,2]$ & $\frac{375975613}{72745944}$ & $\frac{406255085}{72745944}$ &? \\
		\hline \hline
		\multirow{4}{*}{$Sp(8)$} & $[1^{16}]$ & $\frac{305}{24}$ & $\frac{85}{6}$ & Yes; $N_c=10, \ N_f=16$ \\
		\cline{2-5}	
		& $[4^2,2^2,1^4]$ & $\frac{389}{48}$ & $\frac{53}{6}$ &? \\
		\cline{2-5}			
		& $[5^2,3^2]$ & $\frac{30593927}{4642608}$ & $\frac{16735805}{2321304}$ &? \\
		\cline{2-5}		
		& $[5^2,4, 1^2]$ & $\frac{28118905}{4348848}$ & $\frac{3828919}{543606}$ &? \\
		\hline \hline
		\multirow{1}{*}{$Sp(9)$} & $[1^{18}]$ & $\frac{187}{12}$ & $\frac{209}{12}$ & Yes; $N_c=11, \ N_f=18$ \\
		\hline \hline
		$Sp(N-2)$ & $\rho:SU(2) \hookrightarrow Sp(N-2)$ & $a$ & $c$ & 4d $\CN=2$ SUSY \\ 
		\hline \hline
		\multirow{3}{*}{$Sp(9)$} & $[4^2,2^5]$ & $\frac{469}{48}$ & $\frac{509}{48}$ &? \\
		\cline{2-5}		
		& $[6,4,2^3,1^2]$ & $\frac{1426943}{171363}$ & $\frac{12447785}{1370904}$ &? \\
		\cline{2-5}	
		& $[10,4,2^2]$ & $\frac{2035}{384}$ & $\frac{2171}{384}$ &?  \\
		\hline \hline
		\multirow{5}{*}{$Sp(10)$} & $[1^{20}]$ & $\frac{75}{4}$ & $21$ & Yes; $N_c=12, \ N_f=20$ \\
		\cline{2-5}	
		& $[2^3,1^{14}]$ & $\frac{1236439}{70225}$ & $\frac{1398884}{70225}$ &? \\
		\cline{2-5}		
		& $[4,3^2,2^4,1^2]$ & $\frac{5621823}{434312}$ & $\frac{1576769}{108578}$ &? \\
		\cline{2-5}	
		& $[6,4,2, 1^8]$ & $\frac{18160313}{1732800}$ & $\frac{19892353}{1732800}$ &? \\
		\cline{2-5}	
		& $[6,4^2,1^6]$ & $\frac{11747}{1200}$ & $\frac{6371}{600}$ &?  \label{tab:RationalChargesSO}  \\
		\hline												
	\end{longtable}
\end{center}

\section{Generalized Argyres-Douglas theories and $E_6$ SCFT} \label{sec:genAD}
In this section, we apply our general methods to the theories without Lagrangian descriptions. We will be focusing on the theories that satisfies the Sugawara condition between the conformal and flavor central charges \eqref{sugawara}. For the most of the examples we study, we find the $\CN=2$ enhancement occurs only for the principal deformations. But we see some cases exhibiting $\CN=2$ enhancements for the non-principal embeddings as well. There are also a small number of cases where we see rational (but probably $\CN=1$) central charges, but such cases are rare. 

\subsection{$(I_{N, k}, F)$ Argyres-Douglas theory}
Let us first discuss the Argyres-Douglas type theories \cite{Xie:2012hs,Xie:2013jc} with $SU(N)$ flavor symmetry, called the $(I_{N, k}, F)$ theory. When $(N, k)$ are coprime, there is no additional global symmetry other than $SU(N)$. When $N=kn$, this theory has additional global symmetry $U(1)^{N-1}$. The relevant information about this theory has been reviewed in the appendix of \cite{Maruyoshi:2016aim}. 

\paragraph{Principal deformation}
The principal deformation for this theory when $k=Nm+1$ has already been studied in \cite{Maruyoshi:2016aim}. Let us consider the deformations for arbitrary $(N, k)$. We find that for the principal deformation and all $(N, k)$, the deformed theory always flows to the $(A_{N-1}, A_{N+k-1})$ theory:
\begin{align}
 \boxed{\CT_{{\rm IR}} \left[ (I_{N, k}, F) \textrm{ AD theory}, [N] \right]} = \boxed{(A_{N-1}, A_{N-k+1}) \textrm{ AD theory} }
\end{align}

\paragraph{Non-principal deformations}
For most cases, a non-principal embedding does not lead us to an $\CN=2$ fixed point in the IR. But when $k=-N+2$, we find that the embedding given by the partition $[N-1, 1]$ gives rise to the $\CN=2$ supersymmetry in the IR. 
This is the next-to-maximal embedding in the sense described in the introduction.
More precisely, we find that $(I_{N, -N+2}, F)$ theory flows to $(A_1, D_N)$ theory:
\begin{align}
 \boxed{\CT_{{\rm IR}} \left[ (I_{N, -N+2}, F) \textrm{ AD theory}, [N-1, 1] \right]} = \boxed{(A_1, D_N) \textrm{ AD theory} }
\end{align}
When $N$ is even, the $(I_{2n, -2n+2}, F)$ theory is identical to the $SU(n)$ SQCD with $2n$ flavors. We have already discussed the corresponding flow in section \ref{sec:LagA1D2N}. 

What is interesting is when $N=2n+1$ is odd. In this case, we obtain $(A_1, D_{2n+1})$ theory as the end point of the RG flow. But we have seen in section \ref{sec:LagA1D2N+1} that $Sp(n)$ SQCD with $4n+4$ fundamental half-hypermultiplets also flows to the same fixed point upon the nilpotent deformation labelled by $[4n+1, 1^3]$. The latter theory (SQCD) has a Lagrangian description, whereas the former ($(I_{2n+1, -2n+1}, F)$ theory) does not. Therefore we find an IR duality between a Lagrangian theory and a ``non-Lagrangian" $\CN=1$ theory:
\begin{align} 
\begin{split}
\boxed{\CT_{{\rm IR}}[Sp(n) \textrm{ SQCD}, [4n+1, 1^3]]} &= \boxed{\CT_{{\rm IR}}[(I_{2n+1, -2n+1}, F) \textrm{ AD theory}, [2n, 1]]} \\
 &= \boxed{(A_1, D_{2n+1}) \textrm{ AD theory}} 
\end{split}
\end{align}
 For the former, the $SU(2)$ flavor symmetry is visible in the UV, whereas in the latter description, only $U(1)$ symmetry is manifest.

\paragraph{Results}
Let us summarize our results in a table. As it has been the case for all of our previous examples, the trivial embedding ($[1^N]$) does not trigger a non-trivial RG flow. Therefore we recover the original theory (with the decoupled chiral multiplet $M$). 

	\begin{longtable}{|c|c|c|c|c|c|}
		\caption{Nilpotent deformations of the $(I_{N, k}, F)$ theories. We only list the cases that yield rational central charges.} \\ 
		\hline
		$I_{N, k}$ & $\rho:SU(2) \hookrightarrow SU(N)$ & $a$ & $c$ & 4d $\CN=2$ SUSY \\ 
		\hline \hline
		\multirow{1}{*}{$I_{2, 0}$}  & $[2]$ & $\frac{1}{24}$ & $\frac{1}{12}$ & Yes; $(A_1, D_2)$ = 2 free hypers \\
		\hline
		\multirow{1}{*}{$I_{2, 1}$}  & $[2]$ & $\frac{43}{120}$ & $\frac{11}{30}$ & Yes; $(A_1, A_2)$ \\
		\hline
		\multirow{1}{*}{$I_{2, 2}$}  & $[2]$ & $\frac{11}{24}$ & $\frac{1}{2}$ & Yes; $(A_1, A_3)$ \\
		\hline 	
		\multirow{1}{*}{$I_{2, 3}$}  & $[2]$ & $\frac{67}{84}$ & $\frac{17}{21}$ & Yes; $(A_1, A_4)$ \\
		\hline 	
		\multirow{1}{*}{$I_{2, 4}$}  & $[2]$ & $\frac{11}{12}$ & $\frac{23}{24}$ & Yes; $(A_1, A_5)$ \\
		\hline \hline
		\multirow{2}{*}{$I_{3, -1}$} & $[3]$ & $\frac{43}{120}$ & $\frac{11}{30}$ & Yes; $(A_2, A_1)=(A_1, A_2)$ \\
		\cline{2-5} & $[2, 1]$ & $\frac{11}{24}$ & $\frac{1}{2}$ & Yes; $(A_1, D_3)$ \\
		\hline 
		\multirow{2}{*}{$I_{3, 0}$} & $[3]$ & $\frac{7}{12}$ & $\frac{2}{3}$ & Yes; $(A_2, A_2)$ = $(A_1, D_4)$ \\
		\cline{2-5} & $[2, 1]$ & $\frac{3879}{4624}$ & $\frac{2373}{2312}$ & ? \\
		\hline
		\multirow{1}{*}{$I_{3, 1}$} & $[3]$ & $\frac{75}{56}$ & $\frac{19}{14}$ & Yes; $(A_2, A_3)$ \\
		\hline
		\multirow{1}{*}{$I_{3, 2}$} & $[3]$ & $\frac{91}{48}$ & $\frac{23}{12}$ & Yes; $(A_2. A_4)$ \\
		\hline
		\multirow{1}{*}{$I_{3, 3}$} & $[3]$ & $\frac{9}{4}$ &$\frac{7}{3}$ & Yes; $(A_2, A_5)$ \\
		\hline
		\hline \hline	
		\multirow{2}{*}{$I_{4, -2}$} & $[4]$ & $\frac{11}{24}$ & $\frac{1}{2}$ & Yes; $(A_3, A_1)=(A_1, A_3)$  \\ 
		\cline{2-5} & $[3, 1]$ & $\frac{7}{12}$ & $\frac{2}{3}$ & Yes; $(A_1, D_4)$ \\
		\hline
		\multirow{1}{*}{$I_{4, -1}$} & $[1^4]$ & $\frac{75}{56}$ & $\frac{19}{14}$ & Yes; $(A_3, A_2)=(A_2, A_3)$ \\
		\hline
		\multirow{1}{*}{$I_{4, 0}$} & $[1^4]$ & $\frac{15}{8}$ & $2$ & Yes; $(A_3, A_3)$ \\
		\hline
		\multirow{1}{*}{$I_{4, 1}$} & $[1^4]$ & $\frac{115}{36}$ & $\frac{29}{9}$ & Yes; $(A_3, A_4)$ \\
		\hline
		\multirow{1}{*}{$I_{4, 2}$} & $[1^4]$ & $\frac{493}{120}$ & $\frac{25}{6}$ & Yes; $(A_3, A_5)$ \\
		\hline
		\multirow{1}{*}{$I_{4, 3}$} & $[1^4]$ & $\frac{465}{88}$ & $\frac{117}{22}$ & Yes; $(A_3, A_6)$ \\
		\hline \hline
		\multirow{2}{*}{$I_{5, -3}$} & $[5]$ & $\frac{67}{84}$ & $\frac{17}{21}$ & Yes; $(A_4, A_1)=(A_1, A_4)$  \\
		\cline{2-5} & $[4, 1]$ & $\frac{19}{20}$ & $1$ & Yes; $(A_1, D_5)$ \\
		\hline
		\multirow{2}{*}{$I_{5, -2}$} & $[5]$ & $\frac{91}{48}$ & $\frac{23}{12}$ & Yes; $(A_4, A_2)=(A_2, A_4)$ \\
		\cline{2-5} & $[2, 1^3]$ & $\frac{821}{24}$ & $\frac{451}{120}$ & ? \\
		\hline
		\multirow{1}{*}{$I_{5, -1}$} & $[5]$ & $\frac{115}{36}$ & $\frac{29}{9}$& Yes; $(A_4, A_3)=(A_3, A_4)$ \\
		\hline
		\multirow{2}{*}{$I_{5, 0}$} & $[5]$ & $\frac{815}{132}$ & $\frac{205}{33}$& Yes; $(A_4, A_4)$ \\
		\cline{2-5} & $[3, 2]$ & $\frac{1223785}{215472}$ & $\frac{655945}{107736}$ & ? \\
		\hline
		\multirow{1}{*}{$I_{5, 1}$} & $[5]$ & $\frac{187}{24}$ & $\frac{47}{6}$& Yes; $(A_4, A_5)$ \\
		\hline \hline
		\multirow{2}{*}{$I_{6, -4}$} & $[6]$ & $\frac{11}{12}$ & $\frac{23}{24}$& Yes; $(A_5, A_1)=(A_1, A_5)$ \\
		\cline{2-5} & $[5, 1]$ & $\frac{13}{12}$ & $\frac{7}{6}$ & $(A_1, D_6)$ \\
		\hline
		\multirow{2}{*}{$I_{6, -3}$} & $[6]$ & $\frac{9}{4}$ & $\frac{7}{3}$ & Yes; $(A_5, A_2)=(A_2, A_5)$\\
		\cline{2-5} & $[2^2, 1^2]$ & $\frac{6551}{1444}$ & $\frac{1842}{361}$ & ? \\
		\hline
		\multirow{2}{*}{$I_{6, -2}$} & $[6]$ & $\frac{493}{120}$ & $\frac{25}{6}$ & Yes; $(A_5, A_3)=(A_3, A_5)$\\
		\cline{2-5} & $[2, 1^4]$ & $\frac{6021}{784}$ & $\frac{3231}{392}$ & ? \\
		\hline
		\multirow{1}{*}{$I_{6, -1}$} & $[6]$ & $\frac{815}{132}$ & $\frac{205}{33}$ & Yes; $(A_5, A_4)=(A_4, A_5)$\\
		\hline
		\multirow{1}{*}{$I_{6, 0}$} & $[6]$ & $\frac{185}{24}$ & $\frac{95}{12}$ & Yes; $(A_5, A_5)$\\
		\hline
		\multirow{1}{*}{$I_{6, 1}$} & $[6]$ & $\frac{1095}{104}$ & $\frac{275}{26}$ & Yes; $(A_5, A_6)$\\
		\hline
\end{longtable}

\subsection{$E_6$ SCFT} 
Let us study all the nilpotent deformations of the $E_6$ SCFT \cite{Minahan:1996fg}. The relevant information on the nilpotent orbits can be found in the appendix of \cite{Chacaltana:2014jba}. We find that there are only 4 cases which flows to a theory with rational central charges. As before, the trivial embedding gives us the original $E_6$ theory. Three other embeddings trigger flows to $H_0, H_1, H_2$ Argyres-Douglas theories. 
See the table \ref{table:E6} for the summary of the results. 
\begin{table}[!h]
\centering
\begin{tabular}{|c|c|c|c|c|}
	\hline
	$\rho: SU(2) \hookrightarrow E_6$ & $\mathfrak{f}_c$ & $a$ & $c$ & $\CN=2$ theory \\
	\hline \hline
	$E_6$ & $\varnothing$ & $\frac{43}{120}$ & $\frac{11}{30} $& $H_0 = (A_1, A_2)$ \\
	\hline
	$D_5$ & $U(1)$ & $\frac{11}{24}$ & $\frac{1}{2}$ & $H_1 = (A_1, A_3)$ \\
	\hline
	$D_4$ & $SU(3)$ & $\frac{7}{12}$ & $\frac{2}{3}$ & $H_2 = (A_1, D_4)$ \\
	\hline	
\end{tabular}
\caption{Nilpotent Deformations of the $E_6$ SCFT. The nilpotent orbits are given by the Bala-Carter label. The $\mathfrak{f}_c$ denotes the commuting subalgebra under the embedding. The ones not listed in this table trigger flows to $\CN=1$ SCFTs with irrational central charges.}
\label{table:E6}
\end{table}

Combining with the $\CN=1$ theory discovered in \cite{Gadde:2015xta} flowing to the $E_6$ SCFT, it may be possible to write another ``Lagrangian" descriptions for the $H_{0}$, $H_1$, $H_2$ Argyres-Douglas theories. But, the $E_6$ flavor symmetry is not visible in the UV, whereas we have been using the $E_6$ flavor symmetry rather explicitly. It would be interesting to come up with an alternative $\CN=1$ gauge theory flowing to the same AD fixed points.


\section{The full superconformal index of $(A_1, D_{N})$ theory}
\label{sec:SCIA1DN}
 Various limits of the superconformal index for Argyres-Douglas theories were computed in \cite{Buican:2015ina,Cordova:2015nma,Buican:2015tda,Song:2015wta} and the full superconformal index of the $(A_1,A_N)$-type theories was computed in \cite{Maruyoshi:2016tqk,Maruyoshi:2016aim} using the $\CN=1$ RG flow to the AD theories. In this section, we use the Lagrangians of section \ref{sec:LagA1D2N} and \ref{sec:LagA1D2N+1}, to compute the full superconformal indices of the $(A_1, D_{N})$ theories.
 
The superconformal index was first proposed in \cite{Kinney:2005ej,Romelsberger:2005eg}. For an SCFT with $\CN=1$ supersymmetry, it is defined as 
 \be
 \CI_{\CN=1} (p,q,\xi;\vec{a})=\Tr (-1)^F p^{j_1+j_2+\frac{R}{2}} q^{j_2-j_1+\frac{R}{2}}\xi^{\CF}\prod_{i} a_i^{\CF_i} \ .
 \label{eq:SCIN1}
 \ee 
Here, $F$ is the fermion number, $(j_1,j_2)$ are the Cartans of the four-dimensional Lorentz group $SU(2)_1 \times SU(2)_2$, $R$ is the exact $\CN=1$ R-charge at the fixed point, $\CF$ is an axial global symmetry and $\CF_i$ are the Cartans for the global symmetry $H$, acting on the fixed point theory.

The superconformal index gets contributions from the various chiral and vector supermultiplets that go into constructing a given Lagrangian. A chiral multiplet with $\CN=1$ R-charge $R$, $\CF$-charge $f$ and transforming in the representation $\mathfrak{R}$ of $H$, contributes to the index as 
\be
\CI_{\text{chiral}}^{(R, f)}(p,q,\xi;\vec{a}) = \prod_{\vec{w} \in \mathfrak{R}}\Gamma((p q)^{\frac{R}{2}} \xi^{f}\vec{a}^{\vec{w}}) \ . 
\label{eq:SCIchiral}
\ee   
Here, the product runs over all the weight-vectors $\vec{w}$ in the representation $\mathfrak{R}$ of $H$ and $\Gamma(z)$ is the elliptic gamma function defined as
\be
\Gamma(z) \equiv \Gamma(p,q;z) = \prod_{m,n=0}^{\infty}\frac{1- z^{-1} p^{m+1}q^{n+1}}{1-z p^m q^n} \ .
\ee 
We will follow the standard abbreviated notations given by $\vec{a}^{\vec{w}} \equiv \prod_{i} a_i^{w_i}$ and $f(z^{\pm}) \equiv f(z^+) f(z^-)$. When the Lagrangian describes a gauge theory with gauge group $G$, the formula in \eqref{eq:SCIchiral} will also include fugacities with respect to the gauge group. The vector multiplets of $G$ will contribute as 
\be
\CI_{\text{vec}}(p,q)= \kappa^{{\rm rank} (G)} \prod_{\vec{\alpha} \in \Delta_G} \frac{1}{\Gamma(z^{\vec{\alpha}})} \ ,
\label{eq:SCIvec}
\ee  
where $\Delta_G$ is the set of all the roots of $G$ and $\kappa = (p;p)(q;q)$, with $(z;q)$ being the $q$-Pochhammer symbol defined by
\be
(z;q) = \prod_{m=0}^{\infty} (1-zq^m)\ .
\ee 
The full superconformal index is then obtained by taking the product of \eqref{eq:SCIchiral} and \eqref{eq:SCIvec} and integrating over the gauge group. However, if there are gauge invariant operators that hit the unitarity bound along the RG flow and thereby decouple from the interacting theory, then we will have to appropriately remove them in order to obtain the index of the interacting theory. A prescription for this was given in \cite{PANote} and is reproduced in appendix \ref{app:AccidentalIndex}. Also see \cite{Morita:2011cs,Agarwal:2012wd,Safdi:2012re} for a similar prescription to correct the $S^3$ partition function of three-dimensional theories with accidental symmetries.

Similarly, for SCFTs with the $\CN=2$ supersymmetry, the superconformal index is defined by
\be
\CI_{\CN=2}(p,q,t)=\Tr (-1)^{F} p^{j_1+j_2+\frac{r}{2}} q^{j_2-j_1+\frac{r}{2}} t^{R-\frac{r}{2}} \ ,
\label{eq:SCIN2}
\ee 
where $R$ and $r$ are the Cartans of the $SU(2)_R \times U(1)_r$ symmetry of the $\CN=2$ superconformal algebra, with $r$ charge normalized to be such that the exact dimension of Coulomb branch operators is given by $\Delta = \frac{r}{2}$. The fugacities $p,q$ and $t$ are constrained to satisfy 
\be
|p|<1, \ |q|<1, \ |t|<1, \ \left|\frac{p q}{t} \right|<1 \ .
\ee 
For $\CN=2$ SCFTs, we can also compute their $\CN=1$ superconformal index by using \eqref{eq:SCIN1}. This can then be mapped to the $\CN=2$ index by setting $\xi \rightarrow (t (p q)^{-\frac{2}{3}})^\beta$. Here, $\beta$ is determined by the normalization of $U(1)_{\CF}$ inside $SU(2)_R \times U(1)_r$. For our purposes, it will be useful to reparametrize the fugacities such that $p=\mathfrak{t}^3 y, q = \mathfrak{t}^3/y, t = \mathfrak{t}^4/v$. The $\CN=2$ superconformal index then becomes
\be
\CI_{\CN=2}(\mathfrak{t}, y, v)=\Tr (-1)^{F} \mathfrak{t}^{2(E+j_2)}y^{2j_1}v^{-R+\frac{r}{2}} \ ,
\label{eq:SCIN2redef}
\ee 
where, $E$ is the scaling dimension of the operator contributing to the index. The reparametrized index of \eqref{eq:SCIN2redef} can be easily expanded in terms of $\mathfrak{t}$, when doing explicit computations.

\subsection{$(A_1,D_{2N})$ theory}
We can obtain the $\CN=2$ superconformal index of the $(A_1,D_{2N})$ theory by considering the $\CN=1$ index of the theory described in section \ref{sec:LagA1D2N} (with appropriate corrections due to decoupling of operators) and then redefining $\xi \rightarrow (t (p q)^{-\frac{2}{3}})^\beta$, as explained earlier. Modulo decoupled operators, the interacting theory at the IR fixed point is an $SU(N)$ gauge theory with matter fields and their exact ($R_{\CN=1} ,\mathcal{F}$) charges given in table \ref{tab:A1D2N}.
\begin{table}
	\centering
	\begin{tabular}{|c|c|c|c|c|c|}
		\hline
		fields & $SU(N)$ & $U(1)_1$ & $U(1)_2$ & $R_{\CN=1} = \frac{(3N-1)J_+ + J_-}{3N}$ & $\CF = \frac{J_+-  J_-}{2}$\\
		\hline \hline
		$q_1$ & $\mathbf{N}$ & 1 & $2N-1$ &  $\frac{3N-1}{3N}$ &  $\half$ \\ \hline
		$\widetilde{q}_1$ & $\bar{\mathbf{N}}$ & -1 & $-(2N-1)$ &  $\frac{3N-1}{3N}$ &  $\half$\\ \hline
		$q_2$ & $\mathbf{N}$ & 1 & -1 &  $\frac{N+1}{3N}$ & $\frac{2N-1}{2}$ \\ \hline
		$\widetilde{q}_2$ & $\bar{\mathbf{N}}$  & -1 & 1 &  $\frac{N+1}{3N}$  & $\frac{2N-1}{2}$  \\ \hline
		$\phi$ & \rm{adj} & 0 & 0 &  $\frac{2}{3N}$ &  $-1$ \\ \hline
		$M_j$, \tiny{$( N \leq j \leq 2N-2)$} & $\mathbf{1}$ & 0 & 0 & $\frac{2j+2}{3N}$ & $-(j+1)$ \\ \hline
	\end{tabular}
	\caption{Matter content (modulo decoupled Coulomb branch operators) of interacting theory in IR of the ``Lagrangian'' for $(A_1, D_{2N})$ theory.} 
	\label{tab:A1D2N}
\end{table}
The $\CN=1$ superconformal index of this theory is therefore given by
\begin{align}
\label{eq:N1SCIA1D2N}
\CI_{\CN=1}^{(A_1, D_{2N})} &= \frac{\prod_{j=N}^{2N-2} \G \left((pq)^{\frac{j+1}{3N}} \xi^{-(j+1)} \right)}{\prod_{i=2}^N \G \left((pq)^{\frac{i}{3N}} \xi^{-i}\right)}  
 \times \frac{\k^{N-1}}{N!}  \G\left( (pq)^{\frac{1}{3N}} \xi^{-1} \right)^{N-1} \oint [d\vec{z}] \prod_{\vec{\a} \in \Delta} \frac{\G(\vec{z}^{\vec\a} (pq)^{\frac{1}{3N}} \xi^{-1})}{\G(\vec{z}^{\vec\a})} \nn \\
  & \quad \times 
\prod_{\vec{w} \in R} \G \left((\vec{z}^{\vec{w}} a_1 a_2^{2N-1})^{\pm} (pq)^{\frac{3N-1}{6N}} \xi^\half \right) \G \left((\vec{z}^{\vec{w}} a_1 a_2^{-1})^{\pm} (pq)^{\frac{N+1}{6N}} \xi^{\frac{2N-1}{2}} \right) , 
\end{align}	
where $[d\vec{z}] =  \prod_{i=1}^{N-1} \frac{dz_i}{2\pi i z_i}$, $\Delta$ is the set of all non-zero roots of $SU(N)$ and $R$ is the set of weights of the fundamental representation of $SU(N)$, while $a_{1,2}$ are the fugacities for the $U(1)_{1,2}$ global symmetries acting on the theory. The integration contour is given by the unit circle $|z_i|=1$. 

The numerator in the first term of the first line comes from the $M_j$ fields that remain coupled in the IR. The denominator in that term is included in order to account for the decoupling of $\tr \phi^i$ operators. The remaining entries in the above expression come from the gauge fields and matter fields $\phi, q_1, \widetilde{q_1}, q_2$ and  $\widetilde{q_2}$. 

In order to convert \eqref{eq:N1SCIA1D2N} into the $\CN=2$ index of the $(A_1,D_{2N})$ theory, we now redefine $\xi \rightarrow (t(p q)^{-\frac{2}{3}})^{\beta} $ with $\beta =\frac{1}{N}$. This value of $\beta$ can be fixed by using the fact that the exact dimension of the operator $M_j$ is $\frac{j+1}{N}$. Now, since it becomes the Coulomb branch operator of the $(A_1,D_{2N})$ theory, thus upon SUSY enhancement, its $\CN=2$ $R$-charges must be $(R ,r) = (0, 2 \Delta(M_j)) = (0,\frac{2j+2}{N})$. We also expect the $\CN=1$ flavor charge $\CF$ to be related to the $\CN=2$ $R$-charges, such that
\be
\CF=\frac{1}{\beta} (R-\frac{r}{2}) \ .
\label{eq:beta}
\ee 
This is because, from  the point of view of the $\CN=1$ subalgebra of the $\CN=2$ supersymmetry, the theory has a unique axial flavor symmetry which must be proportional to $R-\frac{r}{2}$. Substituting the respective charges of $M_j$, in \eqref{eq:beta}, then gives the result we seek.

We thus find that the $\CN=2$ superconformal index of the $(A_1,D_{2N})$ theory is given by 
\begin{align}
\label{eq:N2SCIA1D2N}
\begin{split}
\CI_{\CN=2}^{(A_1, D_{2N})} &= \frac{\prod_{j=N}^{2N-2} \G \left((\frac{pq}{t})^{\frac{j+1}{N}} \right)}{\prod_{i=2}^N \G \left((\frac{pq}{t})^{\frac{i}{N}} \right)}  
\times \frac{\k^{N-1}}{N!}  \G\left( (\frac{pq}{t})^{\frac{1}{N}} \right)^{N-1} \oint [d\vec{z}] \prod_{\vec{\a} \in \Delta}  \frac{\G(\vec{z}^{\vec\a}(\frac{pq}{t})^{\frac{1}{N}} )}{\G(\vec{z}^{\vec\a})} \\
& \quad \times \prod_{\vec{w} \in R} \G \left((\vec{z}^{\vec{w}} a_1 a_2^{2N-1})^{\pm} (pq)^{\frac{N-1}{2N}} t^{\frac{1}{2N}} \right) \G \left((\vec{z}^{\vec{w}} a_1 a_2^{-1})^{\pm} (pq)^{\frac{1-N}{2N}} t^{\frac{2N-1}{2N}} \right) . 
\end{split}
\end{align}
As explained in \cite{Agarwal:2014rua}, computing the integral in \eqref{eq:N2SCIA1D2N} requires some care. We will therefore, rewrite the fugacities $p,q,t$ in terms of $y,v, \mathfrak{t}$, such that $p=\mathfrak{t}^3 y, q = \mathfrak{t}^3/y, t = \mathfrak{t}^4/v$. This allows us to compute \eqref{eq:N2SCIA1D2N} as a series expansion in $\mathfrak{t}$. 

We expect to obtain the superconformal index for the $(A_1,D_4)$ theory upon substituting $N=2$ in \eqref{eq:N2SCIA1D2N}. Recall that, $(A_1,D_4)$ has an $SU(3)$ flavor symmetry which we expect to emerge as an enhancement of the manifest $U(1)_1 \times U(1)_2$ flavor symmetry of our theory. To see this enhancement in the superconformal index, it will be useful to redefine the $(a_1,a_2) \rightarrow (z_1,z_2)$ such that $z_1 = a_2^2/a_1^{2/3},\ z_2 = a_1^{4/3}$.  The first few terms in the series expansion of the superconformal index are then given by 
\begin{align}
\label{eq:SCIA1D4}
& \CI^{(A_1,D_4)}_{\CN=2} = 1+\mathfrak{t}^3 v^{3/2}+\mathfrak{t} ^4\Big( v^{-1}\chi_{SU(3),\rm{adj}} (z_1,z_2) -\sqrt{v} \chi_{SU(2),f}(y)\Big) +\mathfrak{t}^5v^{-1/2} \nn \\ 
&~+\mathfrak{t}^6 \Big( v^{3/2}\chi_{SU(2),f}(y)+v^3-\chi_{SU(3),\rm{adj}}(z_1,z_2)-1\Big)\nn \\
&~+ \mathfrak{t}^7\Bigg(v^{-1} \chi_{SU(2),f}(y) \Big(\chi_{SU(3),\rm{adj}}(z_1,z_2)+1\Big) -v^2 \chi_{SU(2),f}(y)-v^{1/2}\Big(\chi_{SU(2),\rm{adj}}(y)+1\Big) \Bigg) \nn \\ 
&~+ \mathfrak{t}^8\Bigg(v^{-2}\chi_{SU(3),{\bf{27}}}(z_1,z_2) + 2v + v^{-1/2}\chi_{SU(2),f}(y)\Bigg) \\ 
&~+ \mathfrak{t}^9 \Bigg(v^{9/2}+ v^{3/2} (\chi_{SU(2),\rm{adj}}(y)-1) + v^3 \chi_{SU(2),f}(y)-\chi_{SU(2),f}(y)(\chi_{SU(3),\rm{adj}}(z_1,z_2)+2) \Bigg) \nn \\ 
&~+ \hdots \nn \ ,
\end{align}
where $\chi_{SU(2),\mathfrak{R}}(y)$ denotes the character of the irreducible representation $\mathfrak{R}$ of $SU(2)_1$. Similarly, $\chi_{SU(3),\mathfrak{R}}(z_1,z_2)$ denotes the character of the irreducible representation $\mathfrak{R}$ of $SU(3)$.

Upon substituting $N=3$ in \eqref{eq:N2SCIA1D2N}, we obtain the index of the $(A_1,D_6)$ theory. This time, the $U(1)_1 \times U(1)_2$ global symmetry of the UV gauge theory is expected to enhance to $SU(2)\times U(1)$. In \eqref{eq:A1D2Nflv}, we found the scaling factor required to map the $U(1)_2$ charges to $SU(2)$-spin quantum numbers. This implies that the $SU(2)$ fugacity $z$ is related to $a_2$ by $z=a_2^3$. The first few terms in the series expansion of the index for $(A_1,D_6)$ are 
\begin{align}
\label{eq:SCIA1D6}
I^{(A_1,D_6)}_{\CN=2} &= 1+\mathfrak{t}^{8/3} v^{4/3}+\mathfrak{t}^{10/3} v^{5/3}-\mathfrak{t}^{11/3}v^{1/3} \chi_{SU(2),f}(y)+ \mathfrak{t}^4 v^{-1} \Big(\chi_{SU(2),\rm{adj}}(z)+1\Big) \nn \\
&~ - \mathfrak{t}^{13/3} v^{2/3}\chi_{SU(2),f}(y)+ \mathfrak{t}^{14/3}v^{-2/3}+\mathfrak{t} ^{16/3} \left(v^{8/3}+v^{-1/3}\right)+ \mathfrak{t}^{17/3} v^{4/3} \chi_{SU(2),f}(y) \nn \\
&~+ \mathfrak{t}^6 \Big(v^{-3/2}(a_1^3+\frac{1}{a_1^3})\chi_{SU(2),f}(z) + v^3-\chi_{SU(2),\rm{adj}}(z)-2\Big)\nn \\
&~+ \mathfrak{t} ^{20/3} v^{1/3} \Big(\chi_{SU(2),\rm{adj}}(z)-\chi_{SU(2),\rm{adj}}(y)-1+v^{3}\Big)\nn \\
&~+ \mathfrak{t} ^7 \chi_{SU(2),f}(y) \Big(-2 v^2+2v^{-1}+v^{-1}\chi_{SU(2),\rm{adj}}(z)\Big) +\mathfrak{t} ^{22/3}v^{2/3}\Big(1-\chi_{SU(2),\rm{adj}}(y)\Big) \nn\\
&~-\mathfrak{t} ^{23/3}\Bigg(v^{7/3}\chi_{SU(2),f}(y)+ v^{2/3}\chi_{SU(2),f}(y)\Big(\chi_{SU(2),\rm{adj}}(y)-1\Big)\Bigg) +\mathfrak{t} ^8 (v^4+3v) \nn\\
&~+\mathfrak{t} ^8\Bigg(v\chi_{SU(2),\rm{adj}}(y)+v^{-2}(\chi_{SU(2),{\bf 5}}(z)+ \chi_{SU(2),\rm{adj}}(z)+1)-v^{-1/2}(a_1^3+\frac{1}{a_1^3})\chi_{SU(2),f}(z)\Bigg)\nn\\
&~ +\mathfrak{t} ^{25/3}v^{8/3}\chi_{SU(2),f}(y)\nn\\
&~+\mathfrak{t} ^{26/3}\Bigg(v^{13/3}+v^{4/3}\Big(\chi_{SU(2),\rm{adj}}(y)-\chi_{SU(2),\rm{adj}}(z)\Big) + v^{-5/3}\chi_{SU(2),\rm{adj}}(z) \Bigg)\nn \\
&~+\mathfrak{t} ^9\Bigg(v^3\chi_{SU(2),f}(y)+v^{-3/2} \chi_{SU(2),f}(y)  \chi_{SU(2),f}(z) (a_1^3+\frac{1}{a_1^3})-\chi_{SU(2),f}(y)(\chi_{SU(2),\rm{adj}}(z)+4)\Bigg)\nn\\
&~+ \hdots \ .
\end{align}


Let us consider the Coulomb branch limit of \eqref{eq:N2SCIA1D2N}. This is obtained by letting $p\rightarrow 0, \ q \rightarrow 0, t\rightarrow 0$ while keeping $\frac{p}{q}$ and $\frac{pq}{t}$ fixed. Letting $\frac{pq}{t} = u$, we get
\be
\label{eq:CoulombBranchA1D2N}
\begin{split}
\CI_{\CN=2}^{(A_1,D_{2N})}(u)=\Bigg( \prod_{i=1}^{N-1} \frac{1}{1-u^{\frac{2N-i}{N}}}\Bigg) \times \Bigg[\frac{1}{N!}\prod_{i=1}^{N-1}\frac{1-u^{\frac{i+1}{N}}}{1-u^{\frac{1}{N}}}  \oint[d\vec{z}]\prod_{i\neq j}\frac{1-z_i/z_j}{1-u^{\frac{1}{N}} z_i/z_j}    \Bigg] \ . 
\end{split}
\ee 
Through explicit evaluation for the cases with $N\leq 5$ and $|w|<1$, we were able to check that 
\be
\label{eq:CoulombBranchIdentity}
\prod_{i=1}^{N-1}\frac{1-w^{i+1}}{1-w} \frac{1}{N!} \oint[d\vec{z}]\prod_{i\neq j}\frac{1-z_i/z_j}{1-w z_i/z_j}  = 1 \ .
\ee 
Substituting $w=u^{\frac{1}{N}}$ in the expression on the LHS in \eqref{eq:CoulombBranchIdentity} gives the expression inside the square brackets in \eqref{eq:CoulombBranchA1D2N}. Assuming that the result of \eqref{eq:CoulombBranchIdentity} continues to hold for all $N$, then implies that \eqref{eq:CoulombBranchA1D2N} reduces to 
\be
\label{eq:SimpCoulombA1D2N}
\CI_{\CN=2}^{(A_1,D_{2N})}(u)=\prod_{i=1}^{N-1} \frac{1}{1-u^{\frac{2N-i}{N}}} \ ,
\ee 
 which is exactly what we expect from the Coulomb branch limit of the index of the $(A_1,D_{2N})$ theory. 
 
\subsection{$(A_1,D_{2N+1})$ theory}  As demonstrated in section \ref{sec:LagA1D2N+1}, the fixed point described by the $(A_1,D_{2N+1})$ theory can be reached by following the RG flow of an $\CN=1$ $Sp(N)$ gauge theory with matter content given in table \ref{tab:A1D2N+1}. 
\begin{table}
	\centering
	\begin{tabular}{|c|c|c|c|c|}
		\hline
		fields & $Sp(N)$ & $SO(3)$  & $R_{\CN=1} = \frac{(6N+2)J_+ + J_-}{6N+3} $ & $\CF =  \frac{J_+ - J_-}{2}$\\
		\hline \hline
		$q_1$ & $\mathbf{2N}$ & $\mathbf{3}$  &  $\frac{6N+2}{6N+3}$& $\half$ \\ \hline
		$q_2$ & $\mathbf{2N}$ & $\mathbf{1}$ &   $\frac{2N+2}{6N+3}$& $ \frac{1+4N}{2}$ \\ \hline
		$\phi$ & \rm{adj} & $\mathbf{1}$  & $\frac{2}{6N+3}$ & $-1$\\ \hline
		$M_{j=2k+1}$, \tiny{$(N \leq k \leq 2N-1)$} & $\mathbf{1}$ & $\mathbf{1}$ & $\frac{2j+2}{6N+3}$& $-(j+1)$ \\ \hline
	\end{tabular}
	\caption{Matter content (modulo decoupled Coulomb branch operators) of the interacting $\CN=1$ theory flowing to the $(A_1, D_{2N+1})$ fixed point.} 
	\label{tab:A1D2N+1}
\end{table}
The $\CN=1$ index of this theory is given by
\begin{align}
\label{eq:N1SCIA1D2N+1}
\CI_{\CN=1}^{(A_1, D_{2N+1})} &= \Bigg[\prod_{i=1}^{N}\frac{ \G \left((pq)^{\frac{2N+2i}{6N+3}} \xi^{-(2N+2i)} \right)}{ \G \left((pq)^{\frac{2i}{6N+3}} \xi^{-2i}\right)}\Bigg]  
\times \frac{\k^{N}}{2^N N!}  \G\left( (pq)^{\frac{1}{6N+3}} \xi^{-1} \right)^{N} \nn \\ 
& \quad \times \oint [d\vec{z}] \prod_{\vec{\a} \in \Delta}  \frac{\G(\vec{z}^{\vec\a} (pq)^{\frac{1}{6N+3}} \xi^{-1})}{\G(\vec{z}^{\vec\a})} 
\prod_{\vec{w} \in R} \prod_{\vec{u} \in R'} \G \left(\vec{z}^{\vec{w}} \vec{s}^{\vec{u}} (pq)^{\frac{3N+1}{6N+3}} \xi^\half \right)   \\ 
& \qquad \times \prod_{\vec{w} \in R} \G \left(\vec{z}^{\vec{w}}  (pq)^{\frac{N+1}{6N+3}} \xi^{\frac{4N+1}{2}} \right) , \nn 
\end{align}
where, now, $\Delta$ is the set of all non-zero roots of $Sp(N)$, $R$ is set of all weights in the fundamental representation of $Sp(N)$ and $R'$ is the set of all weights in the vector representation of the $SO(3)$ flavor symmetry. In order to transform this into the corresponding $\CN=2$ superconformal index, we will redefine $\xi$ such that $\xi \rightarrow (t(pq)^{-\frac{2}{3}})^\beta$ with $\beta=\frac{1}{2N+1}$. This implies
\begin{align}
\label{eq:N2SCIA1D2N+1}
\CI_{\CN=2}^{(A_1, D_{2N+1})} &= \Bigg[\prod_{i=1}^{N}\frac{ \G \left((pq)^{\frac{2N+2i}{2N+1}} t^{-\frac{2N+2i}{2N+1}} \right)}{ \G \left((pq)^{\frac{2i}{2N+1}} t^{-\frac{2i}{2N+1}}\right)}\Bigg]  
\times \frac{\k^{N}}{2^N N!}  \G\left( (pq)^{\frac{1}{2N+1}} t^{-\frac{1}{2N+1}} \right)^{N}  \nn \\ 
& \quad \times \oint [d\vec{z}] \prod_{\vec{\a} \in \Delta}  \frac{\G(\vec{z}^{\vec\a} (pq)^{\frac{1}{2N+1}} t^{-\frac{1}{2N+1}})}{\G(\vec{z}^{\vec\a})} 
\prod_{\vec{w} \in R} \prod_{\vec{u} \in R'} \G \left(\vec{z}^{\vec{w}} \vec{s}^{\vec{u}} (pq)^{\frac{N}{2N+1}} t^{\frac{1}{4N+2}} \right)   \\ 
& \qquad \times \prod_{\vec{w} \in R} \G \left(\vec{z}^{\vec{w}}  (pq)^{-\frac{N}{2N+1}} t^{\frac{4N+1}{4N+2}} \right) .\nn
\end{align}
Upon substituting $p=\mathfrak{t}^3 y, q = \mathfrak{t}^3/y, t = \mathfrak{t}^4/v$ in \eqref{eq:N2SCIA1D2N+1},  we can explicitly compute the $\CN=2$ index of the $(A_1,D_{2N+1})$ theory as a series expansion in $\mathfrak{t}$. 

By substituting $N=1$ in \eqref{eq:N2SCIA1D2N+1}, we expect to obtain the superconformal index of $(A_1,D_3)$ theory. The $(A_1,D_3)$ theory is equivalent to the $(A_1,A_3)$ theory, whose full superconformal index was computed in \cite{Maruyoshi:2016aim}. We confirmed that the result obtained from \eqref{eq:N2SCIA1D2N+1}, agrees with that for the $(A_1,A_3)$ theory. Since we will later use this to compute the full superconformal index of the $(A_2,A_5)$ theory, we reproduce the superconformal index of the $(A_1,A_3)$ theory here:
\begin{align}
\label{eq:IndexA1A3}
\begin{split}
\CI_{\CN=2}^{(A_1, A_{3})} &=1+\mathfrak{t} ^{8/3} v^{4/3}-\mathfrak{t} ^{11/3} v^{1/3}\chi_{SU(2),f}(y)+\mathfrak{t} ^4 v^{-1} \chi_{SU(2),\rm{adj}}(z)+\mathfrak{t} ^{14/3}v^{-2/3} \\
&\quad+\mathfrak{t} ^{16/3} v^{8/3}  + \mathfrak{t} ^{17/3} v^{4/3}\chi_{SU(2),f}(y)-\mathfrak{t} ^6 \left(\chi_{SU(2),\rm{adj}}(z)+1\right) \\
&\quad- \mathfrak{t} ^{19/3}v^{5/3}\chi_{SU(2),f}(y)  -\mathfrak{t} ^{20/3}v^{1/3} \left(\chi_{SU(2),\rm{adj}}(y)+1\right) \\
&\quad+\mathfrak{t} ^7v^{-1}\chi_{SU(2),f}(y) \left(\chi_{SU(2),\rm{adj}}(z)+1\right)+\mathfrak{t} ^{22/3} v^{2/3}  +\mathfrak{t} ^{23/3} v^{-2/3}\chi_{SU(2),f}(y)   \\
&\quad +\mathfrak{t} ^8 \left(v^4+v^{-2}\chi_{SU(2),{\bf 5}}(z)+v\right)+ \mathfrak{t} ^{25/3}v^{8/3} \chi_{SU(2),f}(y) \\
&\quad+ \mathfrak{t} ^{26/3}v^{4/3} \left(\chi_{SU(2),\rm{adj}}(y)-1\right) \\
&\quad-  \mathfrak{t} ^9 \Big(v^3 \chi_{SU(2),f}(y)+\big(\chi_{SU(2),\rm{adj}}(z)+2\big)\chi_{SU(2),f}(y)\Big)+ \hdots\ , 
\end{split}
\end{align} 
where $z$ is the fugacity for the $SU(2)$ flavor symmetry with $\chi_{SU(2),\mathfrak{R}}(z)$ being the character for the representation $\mathfrak{R}$ of $SU(2)$.
 
Similarly, we can get the full superconformal indices of the $(A_1,D_5)$ and $(A_1,D_7)$ theories by substituting $N=2,3$ respectively in \eqref{eq:N2SCIA1D2N+1}. We thereby find that
\begin{align}
\label{eq:IndexA1D5}
\begin{split}
\CI^{(A_1,D_5)}_{\CN=2} &= 1+\mathfrak{t} ^{12/5} v^{6/5}+\mathfrak{t} ^{16/5} v^{8/5}-\mathfrak{t} ^{17/5} v^{1/5}\chi_{SU(2),f}(y) \\
&\quad+\mathfrak{t} ^4 v^{-1}\chi_{SO(3),{\bf{3}}}(s)- \mathfrak{t} ^{21/5}v^{3/5}\chi_{SU(2),f}(y)+ \mathfrak{t} ^{22/5}v^{-4/5}+\mathfrak{t} ^{24/5} v^{12/5}  \\
&\quad+\mathfrak{t} ^{26/5}v^{-2/5} +\mathfrak{t} ^{27/5} v^{6/5}\chi_{SU(2),f}(y)+ \mathfrak{t} ^{28/5} v^{14/5}-\mathfrak{t} ^{29/5}v^{7/5} \chi_{SU(2),f}(y)\\
&\quad -\mathfrak{t} ^6\left(\chi_{SO(3),{\bf{3}}}(s)+1\right) + \hdots,
\end{split}
\end{align}
and 
\begin{align}
\label{eq:IndexA1D7}
\begin{split}
\CI^{(A_1,D_7)}_{\CN=2} &= 1+\mathfrak{t} ^4 v^{-1}\chi_{SO(3),{\bf{3}}}(s)+\mathfrak{t} ^{16/7} v^{8/7}+\mathfrak{t} ^{20/7} v^{10/7}-\mathfrak{t} ^{23/7} v^{1/7}\chi_{SU(2),f}(y) \\
&\quad+ \mathfrak{t} ^{24/7} v^{12/7}-\mathfrak{t} ^{27/7} v^{3/7} \chi_{SU(2),f}(y)+\mathfrak{t} ^{30/7}v^{-6/7}-\mathfrak{t} ^{31/7} v^{5/7} \chi_{SU(2),f}(y) \\
&\quad+\mathfrak{t} ^{32/7} v^{16/7} + \mathfrak{t} ^{34/7}v^{-4/7}+\mathfrak{t} ^{36/7} v^{18/7}+\mathfrak{t} ^{37/7} v^{8/7} \chi_{SU(2),f}(y)+\mathfrak{t} ^{38/7}v^{-2/7}\\
&\quad  -\mathfrak{t} ^{39/7} v^{9/7} \chi_{SU(2),f}(y) +2 \mathfrak{t} ^{40/7} v^{20/7}+ \mathfrak{t} ^{41/7} v^{10/7} \chi_{SU(2),f}(y) \\
&\quad -2 \mathfrak{t} ^{43/7} v^{11/7} \chi_{SU(2),f}(y)-\mathfrak{t} ^6(\chi_{SO(3),{\bf 3}}(s)+1)  + \hdots,
\end{split}
\end{align}
where, $\chi_{SO(3),{\bf{3}}}(s)$ is the character of the vector representation of the $SO(3)$ flavor symmetry.

Taking the Coulomb branch limit of \eqref{eq:N2SCIA1D2N+1}, we get
\begin{align}
\label{eq:CoulombBranchA1D2N+1}
\CI_{\CN=2}^{(A_1,D_{2N+1})}(u)=&\Bigg( \prod_{i=1}^{N} \frac{1}{1-u^{2-\frac{2i}{2N+1}}}\Bigg) \times \Bigg[\frac{1}{2^NN!}\prod_{i=1}^{N}\frac{1-u^{\frac{2i}{2N+1}}}{1-u^{\frac{1}{2N+1}}}  \oint[d\vec{z}]\prod_{\vec{\alpha} \in \Delta} \frac{1-\vec{z}^{\vec{\alpha}}}{1-u^{\frac{1}{2N+1}}\vec{z}^{\vec{\alpha}}} \Bigg] 
\end{align}
As before, by explicit computations for $N \leq 5$, we checked that 
\be
\frac{1}{2^NN!}\prod_{i=1}^{N}\frac{1-u^{\frac{2i}{2N+1}}}{1-u^{\frac{1}{2N+1}}}  \oint[d\vec{z}]\prod_{\vec{\alpha} \in \Delta} \frac{1-\vec{z}^{\vec{\alpha}}}{1-u^{\frac{1}{2N+1}}\vec{z}^{\vec{\alpha}}} =1
\ee 
Assuming that the above result continues to hold for all values of $N$, we find that the Coulomb branch limit of \eqref{eq:N2SCIA1D2N+1} is given by 
\begin{align}
\CI_{\CN=2}^{(A_1,D_{2N+1})}(u)=&\Bigg( \prod_{i=1}^{N} \frac{1}{1-u^{2-\frac{2i}{2N+1}}}\Bigg) \ ,
\end{align}
which agrees with the expression we would expect for the Coulomb branch index of $(A_1,D_{2N+1})$ theory.

As another check of our methods leading to \eqref{eq:N2SCIA1D2N} and \eqref{eq:N2SCIA1D2N+1}, we would like to consider its Macdonald and Schur limits \cite{Gadde:2011ik,Gadde:2011uv}. The results for the superconformal index in these limiting
cases have already been obtained, through independent methods, in \cite{Buican:2015ina,Cordova:2015nma,Buican:2015tda,Song:2015wta}. Therefore an agreement between our results and those in existing literature will unequivocally establish the validity of the ``Lagrangians'' for AD theories, being described in this paper.   
 
The Schur limit of an $\CN=2$ superconformal index is obtained by considering $t\rightarrow q$.  As shown in \cite{Gadde:2011uv}, in this limit, the $\CN=2$ index receives a non-zero contribution from only those states which belong to the intersection of cohomologies of two commuting supercharges ($Q_{1+}$ and $\widetilde{Q}_{1\dot -}$ in the notation of \cite{Gadde:2011uv}). This implies that for $\CN=2$ theories, the Schur index is independent of $p$ and can be written as 
\be
\CI_S = \Tr (-1)^F q^{E-R} \ .
\ee
We were not able to find an analytic way of showing that the $p$-dependence of \eqref{eq:N2SCIA1D2N} and \eqref{eq:N2SCIA1D2N+1} drops out in the Schur limit, however, we have checked that this indeed happens for some explicit values of $N$.  For these values of $N$, we also checked that the Schur index obtained from our formula agrees with that given in existing literature. The $p$-independence of the Schur limit of our index hints at the existence of an extra commuting conserved supercharge, which was not present in the $\CN=1$ algebra of the UV theory, and hence, indicates the enhancement of supersymmetry at the IR fixed point of the theories considered here. 

The Macdonald limit of an $\CN=2$ superconformal index is obtained by considering the limit $p\rightarrow0$ while keeping $q$ and $t$ constant. It turns out that the integrand in \eqref{eq:N2SCIA1D2N}, is singular in this limit, so we apply this limit after the integral has been evaluated. In terms of the redefined fugacities $\mathfrak{t},y$ and $v$, this limit requires a rescaling given by
\be
\begin{split}
\mathfrak{t} &\rightarrow \alpha \mathfrak{t} \ , \\
y &\rightarrow \alpha^3 y \ , \\
v &\rightarrow \alpha^4 v \ .
\end{split}
\ee
We can now let $\alpha \rightarrow 0$.  Applying this rescaling to the expressions given in \eqref{eq:SCIA1D4} \eqref{eq:SCIA1D6}, \eqref{eq:IndexA1D5} and \eqref{eq:IndexA1D7} produces the Macdonald indices of the $(A_1,D_4)$, $(A_1,D_6)$, $(A_1,D_5)$ and $(A_1,D_7)$ theories, respectively, as can be checked by comparing with the corresponding results given in \cite{Buican:2015tda, Song:2015wta}.

\subsection{$(A_3,A_3)$ and $(A_2,A_5)$ theory and S-duality} 
As described in \cite{Buican:2014hfa,DelZotto:2015rca,Cecotti:2015hca, Xie:2016uqq}, the $(A_3,A_3)$ theory admits a duality frame where it can be obtained from coupling two copies of $(A_1,D_4)$ theory by gauging an $SU(2)$ subgroup of their diagonal $SU(3)$ flavor symmetry. In addition to this, there is also a doublet of hypermultiplets coupled to the $SU(2)$ gauge group. The whole setup is such that the $SU(2)$ gauge coupling is exactly marginal. In our formulation of the Lagrangian of the $(A_1,D_4)$ theory, the $SU(3)$ flavor symmetry is emergent with only its Cartan subgroup being manifest in the UV. This makes it hard to implement the above mentioned procedure at the level of the Lagrangian, to obtain a Lagrangian for the $(A_3,A_3)$ theory. Nonetheless, we can use this duality frame to obtain the full superconformal index of the $(A_3,A_3)$ theory. This is given by 
\begin{align}
\label{eq:SCIA3A3}
\CI^{(A_3,A_3)}_{\CN=2} = \frac{\kappa}{2!} \oint \frac{dw}{2\pi i w} \frac{\Gamma(\frac{pq}{t}w^{\pm 2, 0}) \Gamma(t^{1/2} w^{\pm 1} x_1^{\pm 1})}{\Gamma(w^{\pm 2})} \CI^{(A_1,D_4)}_{\CN=2} (w x_2^{\frac{1}{3}}, x_2^{\frac{2}{3}})\CI^{(A_1,D_4)}_{\CN=2}(w x_2^{\frac{1}{3}}, x_2^{\frac{2}{3}}) ,
\end{align}
where the repeated sign means taking products with each sign and $w$ is the fugacity for the $SU(2)$ symmetry we gauge and $\CI^{(A_1,D_4)}_{\CN=2}(w x^{\frac{1}{3}}, x^{\frac{2}{3}})$ is superconformal index of the $(A_1,D_4)$ theory given in \eqref{eq:SCIA1D4} with $x$ being the fugacity for the $U(1)$ commutant of $SU(2) \subset SU(3)$. The contribution of the hypermultiplet is encapsulated in the term $\Gamma(t^{1/2} z^{\pm 1} x_1^{\pm 1})$ with $x_1$ being the fugacity for the $SO(2)$ flavor symmetry rotating the two half-hypers in to each other. The remaining pieces in \eqref{eq:SCIA3A3} come from the contribution of the $\CN=2$ vector multiplet for the $SU(2)$ gauge group and $SU(2)$ Haar measure. Explicitly, the index in \eqref{eq:SCIA3A3} takes the form
\begin{align}
\label{eq:ExplicitSCIA3A3}
\begin{split}
& \CI^{(A_3,A_3)}_{\CN=2}(\vec{x}) =1+2 \mathfrak{t} ^3 v^{3/2}+\mathfrak{t} ^4 \left(v^2-2 v^{1/2} \chi_{SU(2),f}(y)+3 v^{-1}\right)+\mathfrak{t} ^5 \left(-v \chi_{SU(2),f}(y)+2v^{-1/2}\right) \\
&~+ \mathfrak{t} ^6 \left(v^{-3/2} (x_1 x_2+\frac{x_2}{x_1}+x_1 x_3+\frac{x_3}{ x_1}+\frac{x_1}{x_2}+\frac{1}{x_1 x_2}+\frac{x_1}{x_3}+\frac{1}{ x_1 x_3})+2 v^{3/2} \chi_{SU(2),f}(y)+3 v^3-3\right)\\
&~+ \mathfrak{t} ^7 \left(2 v^{7/2}-3 v^2 \chi_{SU(2),f}(y)-2 v^{1/2} (\chi_{SU(2),\rm{adj}}(y)-1) +\frac{4}{v} \chi_{SU(2),f}(y)\right) \\
&~+ \mathfrak{t} ^8 \Big(-v^{-1/2}\left(x_1 x_2 + x_1 x_3 +\frac{x_2}{x_1}+\frac{x_3}{x_1}+\frac{x_1}{x_2}+\frac{1}{x_1 x_2}+\frac{x_1}{ x_3}+\frac{1}{ x_1 x_3}+2 \chi_{SU(2),f}(y)\right)  \\
&~ \hspace{1cm}+ v^4+v^{-2}(x_1^2+x_2 x_3+\frac{x_3}{x_2}+\frac{1}{x_1^2}+\frac{x_2}{x_3}+\frac{1}{x_2 x_3}+6)-4 v^{5/2} \chi_{SU(2),f}(y)+7 v\Big) + \hdots \ ,
\end{split}
\end{align}
where $\vec{x} = (x_1,x_2,x_3)$. In \cite{Buican:2014hfa, Buican:2015ina, Buican:2015tda}, it was pointed out that the $(A_3,A_3)$ theory admits an S-duality transformation that permutes the generators of the $U(1)^3$ flavor symmetry group. This transformation mixes the corresponding fugacities $(x_1,x_2,x_3)$, such that
\be
\label{eq:StransA3A3}
x_1 \rightarrow \sqrt{x_2/x_3}, \quad x_2 \rightarrow x_1 \sqrt{ x_2 x_3}, \quad x_3 \rightarrow \frac{\sqrt{x_2 x_3}}{x_1} \ .
\ee  
Invariance of $(A_3,A_3)$ under S-duality then implies that its superconformal index must also be invariant under the transformation given by \eqref{eq:StransA3A3}. It is straightforward to check that this is indeed the case for the expression given in \eqref{eq:ExplicitSCIA3A3}. This further strengthens our confidence that the expression for the full superconformal index of $(A_1,D_4)$ is given by \eqref{eq:SCIA1D4}.

There also exists a duality frame in which the $(A_2,A_5)$ theory can be obtained by coupling an $(A_1,A_3)$ theory with $(A_1,D_6)$ theory and a doublet of hypermultiplets, by gauging their diagonal $SU(2)$ flavor symmetry. As before, the $SU(2)$ flavor symmetry of these systems is an emergent symmetry from the point of view of our Lagrangians and hence, it not clear how to implement the above procedure at the level of the Lagrangians. However, the superconformal index of the $(A_2, A_5)$ theory is straight forward to compute, using this data. It is given by 
\begin{align}
\label{eq:SCIA2A5}
\CI^{(A_2,A_5)}_{\CN=2} = \frac{\kappa}{2!} \oint \frac{dw}{2\pi i w} \frac{\Gamma(\frac{pq}{t}w^{\pm 2, 0}) \Gamma(t^{1/2} w^{\pm 1} x_1^{\pm 1})}{\Gamma(w^{\pm 2})} \CI^{(A_1,A_3)}_{\CN=2} (w)\CI^{(A_1,D_6)}_{\CN=2}(x_2^{\frac{1}{3}},w) \ ,
\end{align}
where $\CI^{(A_1,A_3)}_{\CN=2}(w)$ is the index of the $(A_1,A_3)$ theory as given in \eqref{eq:IndexA1A3} with $w$ here playing the role of $z$ there. $\CI^{A_1,D_6}_{\CN=2}(x_2^{\frac{1}{3}},w)$ is the index of the $(A_1,D_6)$ theory given in \eqref{eq:SCIA1D6}, with $(x_2^{\frac{1}{3}},w)$ here, playing the role of $(a_1,z)$ there.  Explicitly, the first few terms in the index of the $(A_2,A_5)$ theory are found to be

\begin{align}
\label{eq:IndexA2A5}
\CI^{(A_2,A_5)}_{\CN=2} &= 1+2 \mathfrak{t} ^{8/3} v^{4/3}+\mathfrak{t} ^{10/3} v^{5/3}-2v^{1/3}\mathfrak{t} ^{11/3} \chi_{SU(2),f}(y)+\mathfrak{t} ^4 \left(v^2+2 v^{-1}\right)-\mathfrak{t} ^{13/3} v^{2/3}\chi_{SU(2),f}(y) \nn\\
&~+  2 \mathfrak{t} ^{14/3} v^{-2/3}-\mathfrak{t} ^5 v \ \chi_{SU(2),f}(y) + \mathfrak{t} ^{16/3} \left(3 v^{8/3}+v^{-1/3}\right) +2\mathfrak{t} ^{17/3}v^{4/3} \chi_{SU(2),f}(y) \nn\\
&~+\ 2\mathfrak{t} ^6 \left( v^3-1\right)-3 \mathfrak{t} ^{19/3}v^{5/3} \chi_{SU(2),f}(y) + \mathfrak{t}^{20/3}\Bigg(3v^{10/3}-v^{1/3}\Big(2\chi_{SU(2),\rm{adj}}(y)-1\Big) \Bigg)\nn\\
&~+3\mathfrak{t} ^7\Big(v^{-1}-v^2\Big) \chi_{SU(2),f}(y)+\mathfrak{t} ^{22/3} \left(6 v^{2/3}+v^{11/3}\right) -\mathfrak{t} ^{23/3}(v^{-2/3}+5v^{7/3})\chi_{SU(2),f}(y)\nn\\
&~+\mathfrak{t} ^8 \left(5 v^4+\frac{x_1^2}{v^2}+\frac{x_1 x_2}{v^2}+\frac{x_2}{v^2 x_1}+\frac{1}{v^2 x_1^2}+\frac{x_1}{v^2 x_2}+\frac{1}{v^2 x_1 x_2}+\frac{3}{v^2}+ v\Big(\chi_{SU(2),\rm{adj}}(y)+6\Big)\right)\nn\\
&~+\mathfrak{t} ^{25/3} \Big(2v^{8/3}-3v^{-1/3}\Big)\chi_{SU(2),f}(y) +\mathfrak{t} ^{26/3} \Bigg(v^{4/3}\Big(4 \chi_{SU(2),\rm{adj}}(y) +1\Big)+3 v^{13/3}+3v^{-5/3}\Bigg)\nn\\
&~-\mathfrak{t} ^9\Big(7+3v^3\Big) \chi_{SU(2),f}(y)
+\hdots \ ,
\end{align}


The $(A_2,A_5)$ theory is expected to be invariant under a duality group that is isomorphic to $S_3$ \cite{Cecotti:2015hca}. If we denote the generators of this $S_3$ by $f$ and $g$ such that 
\be
f^3=g^2=(fg)^2=1 \ ,
\ee
then the action of $f$ and $g$ on the $U(1)^2$ flavor fugacities, $x_1$,  $x_2$, is given by 
\be
\begin{split}
&f: x_1 \rightarrow \frac{1}{\sqrt{x_1 x_2}} , \ x_2 \rightarrow \sqrt{\frac{x_1^3}{x_2}} \\
&g:x_1 \rightarrow \frac{1}{x_1}, \ x_2\rightarrow x_2 \ .
\end{split}
\ee
It is straightforward to check that the expression given in \eqref{eq:IndexA2A5} is invariant under the $S_3$ duality group of the $(A_2,A_5)$  theory. This provides another non-trivial check for the validity of out proposal for the superconformal indices of the $(A_1,A_3)$ and $(A_1,D_6)$ theories respectively.



\acknowledgments
We would like to thank Philip Argyres, Eduardo Conde, Amihay Hanany, Ken Intriligator, Rudolph Kalveks, Emily Nardoni, Sridip Pal, Fidel Schaposnik, Yuji Tachikawa, Anderson Trimm and Junya Yagi for helpful discussions. 
PA and JS would also like to thank the hospitality of the Simons Center for Geometry and Physics during the 2016 Summer Workshop in Mathematics and Physics.
KM would like to thank the hospitality of ICTP, Trieste and Imperial College London.
JS would also like to thank the hospitality of the Korea Institute for Advanced Study. 
The work of PA is supported by Samsung Science and Technology Foundation under Project Number SSTF-BA1402-08.
The work of JS is supported in part by the US Department of Energy under UCSD's contract de-sc0009919 and also by Hwa-Ahm foundation.

\appendix
\section{Accidental symmetries and superconformal index of adjoint SQCD}
\label{app:AccidentalIndex}

Let us consider a supersymmetric $SU(N_c)$ gauge theory with $N_f$ fundamentals $Q$, $N_f$ anti-fundamentals $\widetilde{Q}$ and an adjoint field $X$. Let the superpotential of our theory be
\be
W=\Tr X^{k+1} \ .
\label{eq:suppotElec}
\ee
This theory is asymptotically free when $N_f < 2 N_c$. A dual theory for the case when $k=2$  was proposed in \cite{Kutasov:1995ve} while the cases with more general $k$ were studied in \cite{Kutasov:1995np}. For our purposes, we will choose $k=2$. The $R$-charges are then given by
\be
\begin{split}
R_{Q} = R_{\widetilde{Q}} &= 1-\frac{2N_c}{3N_f} \ , \\
R_X=\frac{2}{3} \ .
\end{split}
\label{eq:RElec}
\ee 

The magnetic dual to the above theory is given by an $SU(2N_f-N_c)$ gauge theory with $N_f$ fundamentals and anti-fundamentals, $q$, $\widetilde{q}$ along, an adjoint chiral field $Y$ and two chiral singlet fields $M$ and $N$ transforming in the bifundamental of the $SU(N_f)_L \times SU(N_f)_R$ flavor symmetry. Here, $M$ is dual to $Q \widetilde{Q}$ and $N$ dual to $Q X \widetilde{Q}$ in the electric theory. The superpotential of the magnetic theory is given by 
\be
W=\Tr M q Y \widetilde{q} + \Tr N q \widetilde{q} + \Tr Y^3 \ .
\label{eq:suppotMag}
\ee
The corresponding R-charges of the above fields are
\be
\begin{split}
R_q = R_{\widetilde{q}} & = 1-\frac{2}{3}\frac{2N_f-N_c}{N_f} \ , \\
                             R_Y & = \frac{2}{3} \ , \\
                             R_M & = 2-\frac{4 N_c}{3 N_f} \ , \\
                             R_N & = \frac{8}{3} - \frac{4 N_c}{3 N_f} \ .
\end{split}
\label{eq:RMag}
\ee
Once again, the magnetic theory is asymptotically free when $N_f < 2\widetilde{N_c}$ i.e. $N_f > \frac{2}{3} N_c$, where $\widetilde{N_c} = 2N_f -N_c$. Requiring that both electric and magnetic theories be asymptotically free then restricts the range of $N_f$ to be $\frac{2}{3} N_c < N_f < 2 N_c$. In what follows, we will only consider $N_f$ to be in this range.

As was explained in \cite{Kutasov:1995ve}, the effect of the terms $\Tr M q Y \widetilde{q} $ and $\Tr N q \widetilde{q}$ on the magnetic superpotential \eqref{eq:suppotMag}, can be understood by first considering the theory with a superpotential in which these terms are absent. This is then same as the electric theory but with $N_c \rightarrow 2N_f-N_c$. We can now ask: For what values of $N_f$, do the operators $\Tr M q Y \widetilde{q} $  and $\Tr N q \widetilde{q}$ give a relevant deformation of the IR fixed point? It is a simple exercise to check that the operator $\Tr M q Y \widetilde{q} $ is relevant when $N_f > N_c$. When $N_f<N_c$, the dimensions of $M$, as calculated from \eqref{eq:RMag}, violate the unitarity bound and hence, for these values of $N_f$, $M$ remains decoupled from the rest of the theory.  Similar analysis reveals that $\Tr N q \widetilde{q}$ is a relevant deformation when $N_f > \frac{2}{3} N_c$ but becomes irrelevant when $N_f < \frac{2}{3} N_c$, causing $N$ to stay decoupled for the latter values of $N_f$. The corresponding story on the electric side is also quite simple to reproduce. When $N_f < N_c$, The dimensions of the operator $ Q \widetilde{Q}$ violates the unitarity bound causing it to decouple as a free field from the rest of the theory. This replicates the decoupling of $M$ on the magnetic side. Similarly, when $N_f < \frac{2}{3} N_c$, the dimensions of the operator $Q X \widetilde{Q}$ violates the unitarity bound causing it to become a decoupled free field, replicating the decoupling of $N$ in the magnetic theory.

We now want to use the above set up to study how the decoupling of operators affects the superconformal index of a theory. Before incorporating any corrections due to decoupled operators the superconformal index of the electric theory is given by 
\be
\begin{split}
I_E &=  \frac{(p;p)^{N_c-1}_\infty (q;q)^{N_c-1}_\infty }{N_c!} \Gamma(U;p,q)^{N_c-1} \int_{\mathbb{T}^{N_c-1}}  \prod_{j=1}^{N_c-1} \frac{d z_j}{2 \pi i z_j} \\
&~~\times \prod_{1\leq i < j \leq N_c} \frac{\Gamma(U z_i z_j^{-1}, U z_i^{-1}z_j;p,q)}{\Gamma(z_iz_j^{-1},z_i^{-1}z_j;p,q)} 
           \prod_{i=1}^{N_f} \prod_{j=1}^{N_c} \Gamma(s_iz_j, t_i^{-1}z_j^{-1};p,q) \ .
\end{split}
\label{eq:IndexElectricUncorr}
\ee
Here, $z_i$ are fugacities for the $SU(N_c)$ gauge group and are constrained to be such that $\prod_{j=1}^{N_c} z_j=1$. Similarly, $s_i$ and $t_i$ are fugacities for the $SU(N_f)_{L,R}$ flavor groups respectively and $U=(p q)^{\frac{1}{3}}$.    

Before taking into account the decoupling of fields $M$ and $N$, the superconformal index of the magnetic theory is given by
\be
\begin{split}
I_M &= \frac{(p;p)^{\widetilde{N_c}-1}_\infty (q;q)^{\widetilde{N_c}-1}_\infty }{N_c!} \Gamma(U;p,q)^{\widetilde{N_c}-1} \prod_{1\leq i,j \leq N_f}\Gamma (s_i t_j^{-1};p,q) \prod_{1\leq i,j \leq N_f}\Gamma (U s_i t_j^{-1};p,q)\\
   &\quad \times  \int_{\mathbb{T}^{\widetilde{N_c}-1}}  \prod_{j=1}^{\widetilde{N_c}-1} \frac{d z_j}{2 \pi i z_j} \prod_{1\leq i < j \leq \widetilde{N_c}} \frac{\Gamma(U z_i z_j^{-1}, U z_i^{-1}z_j;p,q)}{\Gamma(z_iz_j^{-1},z_i^{-1}z_j;p,q)} \\
 &~ \qquad \times  \prod_{i=1}^{N_f} \prod_{j=1}^{\widetilde{N_c}} \Gamma((UST)^{\frac{1}{\widetilde{N_c}}}s_i^{-1}z_j, (UST)^{-\frac{1}{\widetilde{N_c}}}t_i z_j^{-1};p,q) \ .
\end{split}
\label{eq:IndexMagUncorr}
\ee
Here $S = \prod_{j=1}^{N_c} s_j=1$ and $T = \prod_{j=1}^{N_c} t_j=1$. The term $ \prod_{1\leq i,j \leq N_f}\Gamma (s_i t_j^{-1};p,q)$ represents the contribution of the singlets $M$ to the index while the contribution of the field $N$ is captured by the term $ \prod_{1\leq i,j \leq N_f}\Gamma (U s_i t_j^{-1};p,q)$.

When $N_f> N_c$, the above indices need no correction. The equality of $I_E$ and $I_M$ was verified in the large-$N_c$ limit in \cite{Dolan:2008qi}. Some necessary conditions required for their equality at finite-$N_c$ were proposed and verified in \cite{Spiridonov:2009za}. We will hereon assume that $I_E$ and $I_M$ are equal are finite $N_c$.

Let us now consider the situation when $N_f<N_c$. As was mentioned earlier, for these values of $N_f$, the operator $Q \widetilde{Q}$, in the electric theory, decouples. The index of the interacting theory therefore needs to be corrected in order to account for this.  Taking clue from \cite{Kutasov:2003iy}, we factor out the contribution of $Q \widetilde{Q}$  from $I_E$ and claim that this will then give us the index of the interacting electric theory when $N_f<N_c$. We will verify our claim by comparing the index so obtained with that of the interacting magnetic theory. The corrected electric index thus becomes 
\be
I_{E, N_f<N_c} = \frac{I_E}{ \prod_{1\leq i,j \leq N_f}\Gamma ( s_i t_j^{-1};p,q)} \ .
\label{eq:IndexElecCorr}
\ee

On the magnetic side, for $N_f <N_c$, the field $M$ remains free. Its contribution should therefore not be included in the magnetic SCI for the interacting theory. The corrected magnetic index therefore becomes
\be
I_{M, N_f<N_c} = \frac{I_M}{ \prod_{1\leq i,j \leq N_f}\Gamma ( s_i t_j^{-1};p,q)} \ .
\label{eq:IndexMagCorr}
\ee 
 The equality of $I_{E, N_f<N_c} $ and $I_{M, N_f<N_c} $ follows trivially from the equality of $I_E$ and $I_M$, thus verifying our claim that $I_{E, N_f<N_c} $ as given in \eqref{eq:IndexElecCorr}, represents the index of the interacting theory on the electric side.



\bibliographystyle{jhep}
\bibliography{ADN1}

\providecommand{\href}[2]{#2}\begingroup\raggedright\begin{thebibliography}{10}

\bibitem{Maruyoshi:2016aim}
K.~Maruyoshi and J.~Song, \emph{{N=1 Deformations and RG Flows of N=2 SCFTs}},
  \href{http://arxiv.org/abs/1607.04281}{{\tt 1607.04281}}.

\bibitem{Seiberg:1994bp}
N.~Seiberg, \emph{{The Power of holomorphy: Exact results in 4-D SUSY field
  theories}},  in \emph{{PASCOS '94: Proceedings, 4th International Symposium
  on Particles, Strings and Cosmology, Syracuse, New York, USA, May 19-24,
  1994}}, pp.~0357--369, 1994.
\newblock \href{http://arxiv.org/abs/hep-th/9408013}{{\tt hep-th/9408013}}.

\bibitem{Gadde:2013fma}
A.~Gadde, K.~Maruyoshi, Y.~Tachikawa and W.~Yan, \emph{{New N=1 Dualities}},
  \href{http://dx.doi.org/10.1007/JHEP06(2013)056}{\emph{JHEP} {\bf 1306}
  (2013) 056}, [\href{http://arxiv.org/abs/1303.0836}{{\tt 1303.0836}}].

\bibitem{Agarwal:2013uga}
P.~Agarwal and J.~Song, \emph{{New N=1 Dualities from M5-branes and
  Outer-automorphism Twists}},
  \href{http://dx.doi.org/10.1007/JHEP03(2014)133}{\emph{JHEP} {\bf 1403}
  (2014) 133}, [\href{http://arxiv.org/abs/1311.2945}{{\tt 1311.2945}}].

\bibitem{Agarwal:2014rua}
P.~Agarwal, I.~Bah, K.~Maruyoshi and J.~Song, \emph{{Quiver tails and $
  \mathcal{N}=1 $ SCFTs from M5-branes}},
  \href{http://dx.doi.org/10.1007/JHEP03(2015)049}{\emph{JHEP} {\bf 1503}
  (2015) 049}, [\href{http://arxiv.org/abs/1409.1908}{{\tt 1409.1908}}].

\bibitem{Agarwal:2015vla}
P.~Agarwal, K.~Intriligator and J.~Song, \emph{{Infinitely many $ \mathcal{N}=1
  $ dualities from m + 1 － m = 1}},
  \href{http://dx.doi.org/10.1007/JHEP10(2015)035}{\emph{JHEP} {\bf 10} (2015)
  035}, [\href{http://arxiv.org/abs/1505.00255}{{\tt 1505.00255}}].

\bibitem{Fazzi:2016eec}
M.~Fazzi and S.~Giacomelli, \emph{{$\mathcal{N} = 1$ superconformal theories
  with $D_N$ blocks}},  \href{http://arxiv.org/abs/1609.08156}{{\tt
  1609.08156}}.

\bibitem{Maruyoshi:2016tqk}
K.~Maruyoshi and J.~Song, \emph{{The Full Superconformal Index of the
  Argyres-Douglas Theory}},  \href{http://arxiv.org/abs/1606.05632}{{\tt
  1606.05632}}.

\bibitem{Intriligator:2003jj}
K.~A. Intriligator and B.~Wecht, \emph{{The Exact superconformal R symmetry
  maximizes a}},
  \href{http://dx.doi.org/10.1016/S0550-3213(03)00459-0}{\emph{Nucl. Phys.}
  {\bf B667} (2003) 183--200}, [\href{http://arxiv.org/abs/hep-th/0304128}{{\tt
  hep-th/0304128}}].

\bibitem{Kutasov:2003iy}
D.~Kutasov, A.~Parnachev and D.~A. Sahakyan, \emph{{Central charges and U(1)(R)
  symmetries in N=1 superYang-Mills}},
  \href{http://dx.doi.org/10.1088/1126-6708/2003/11/013}{\emph{JHEP} {\bf 11}
  (2003) 013}, [\href{http://arxiv.org/abs/hep-th/0308071}{{\tt
  hep-th/0308071}}].

\bibitem{Argyres:1995jj}
P.~C. Argyres and M.~R. Douglas, \emph{{New phenomena in SU(3) supersymmetric
  gauge theory}},
  \href{http://dx.doi.org/10.1016/0550-3213(95)00281-V}{\emph{Nucl. Phys.} {\bf
  B448} (1995) 93--126}, [\href{http://arxiv.org/abs/hep-th/9505062}{{\tt
  hep-th/9505062}}].

\bibitem{Argyres:1995xn}
P.~C. Argyres, M.~R. Plesser, N.~Seiberg and E.~Witten, \emph{{New N=2
  superconformal field theories in four-dimensions}},
  \href{http://dx.doi.org/10.1016/0550-3213(95)00671-0}{\emph{Nucl. Phys.} {\bf
  B461} (1996) 71--84}, [\href{http://arxiv.org/abs/hep-th/9511154}{{\tt
  hep-th/9511154}}].

\bibitem{Eguchi:1996vu}
T.~Eguchi, K.~Hori, K.~Ito and S.-K. Yang, \emph{{Study of N=2 superconformal
  field theories in four-dimensions}},
  \href{http://dx.doi.org/10.1016/0550-3213(96)00188-5}{\emph{Nucl. Phys.} {\bf
  B471} (1996) 430--444}, [\href{http://arxiv.org/abs/hep-th/9603002}{{\tt
  hep-th/9603002}}].

\bibitem{Eguchi:1996ds}
T.~Eguchi and K.~Hori, \emph{{N=2 superconformal field theories in
  four-dimensions and A-D-E classification}},  in \emph{{The mathematical
  beauty of physics: A memorial volume for Claude Itzykson. Proceedings,
  Conference, Saclay, France, June 5-7, 1996}}, pp.~67--82, 1996.
\newblock \href{http://arxiv.org/abs/hep-th/9607125}{{\tt hep-th/9607125}}.

\bibitem{Gaiotto:2009hg}
D.~Gaiotto, G.~W. Moore and A.~Neitzke, \emph{{Wall-crossing, Hitchin Systems,
  and the WKB Approximation}},  \href{http://arxiv.org/abs/0907.3987}{{\tt
  0907.3987}}.

\bibitem{Bonelli:2011aa}
G.~Bonelli, K.~Maruyoshi and A.~Tanzini, \emph{{Wild Quiver Gauge Theories}},
  \href{http://dx.doi.org/10.1007/JHEP02(2012)031}{\emph{JHEP} {\bf 1202}
  (2012) 031}, [\href{http://arxiv.org/abs/1112.1691}{{\tt 1112.1691}}].

\bibitem{Gaiotto:2012sf}
D.~Gaiotto and J.~Teschner, \emph{{Irregular singularities in Liouville theory
  and Argyres-Douglas type gauge theories, I}},
  \href{http://dx.doi.org/10.1007/JHEP12(2012)050}{\emph{JHEP} {\bf 12} (2012)
  050}, [\href{http://arxiv.org/abs/1203.1052}{{\tt 1203.1052}}].

\bibitem{Xie:2012hs}
D.~Xie, \emph{{General Argyres-Douglas Theory}},
  \href{http://dx.doi.org/10.1007/JHEP01(2013)100}{\emph{JHEP} {\bf 1301}
  (2013) 100}, [\href{http://arxiv.org/abs/1204.2270}{{\tt 1204.2270}}].

\bibitem{Aharony:2007dj}
O.~Aharony and Y.~Tachikawa, \emph{{A Holographic computation of the central
  charges of d=4, N=2 SCFTs}},
  \href{http://dx.doi.org/10.1088/1126-6708/2008/01/037}{\emph{JHEP} {\bf 01}
  (2008) 037}, [\href{http://arxiv.org/abs/0711.4532}{{\tt 0711.4532}}].

\bibitem{Shapere:2008zf}
A.~D. Shapere and Y.~Tachikawa, \emph{{Central charges of N=2 superconformal
  field theories in four dimensions}},
  \href{http://dx.doi.org/10.1088/1126-6708/2008/09/109}{\emph{JHEP} {\bf 09}
  (2008) 109}, [\href{http://arxiv.org/abs/0804.1957}{{\tt 0804.1957}}].

\bibitem{Xie:2013jc}
D.~Xie and P.~Zhao, \emph{{Central charges and RG flow of strongly-coupled N=2
  theory}}, \href{http://dx.doi.org/10.1007/JHEP03(2013)006}{\emph{JHEP} {\bf
  03} (2013) 006}, [\href{http://arxiv.org/abs/1301.0210}{{\tt 1301.0210}}].

\bibitem{Shapere:1999xr}
A.~D. Shapere and C.~Vafa, \emph{{BPS structure of Argyres-Douglas
  superconformal theories}},  \href{http://arxiv.org/abs/hep-th/9910182}{{\tt
  hep-th/9910182}}.

\bibitem{Cecotti:2010fi}
S.~Cecotti, A.~Neitzke and C.~Vafa, \emph{{R-Twisting and 4d/2d
  Correspondences}},  \href{http://arxiv.org/abs/1006.3435}{{\tt 1006.3435}}.

\bibitem{Cecotti:2011rv}
S.~Cecotti and C.~Vafa, \emph{{Classification of complete N=2 supersymmetric
  theories in 4 dimensions}}, {\emph{Surveys in differential geometry} {\bf 18}
  (2013) }, [\href{http://arxiv.org/abs/1103.5832}{{\tt 1103.5832}}].

\bibitem{Alim:2011ae}
M.~Alim, S.~Cecotti, C.~Cordova, S.~Espahbodi, A.~Rastogi and C.~Vafa,
  \emph{{BPS Quivers and Spectra of Complete N=2 Quantum Field Theories}},
  \href{http://dx.doi.org/10.1007/s00220-013-1789-8}{\emph{Commun. Math. Phys.}
  {\bf 323} (2013) 1185--1227}, [\href{http://arxiv.org/abs/1109.4941}{{\tt
  1109.4941}}].

\bibitem{Alim:2011kw}
M.~Alim, S.~Cecotti, C.~Cordova, S.~Espahbodi, A.~Rastogi and C.~Vafa,
  \emph{{$\mathcal{N} = 2$ quantum field theories and their BPS quivers}},
  \href{http://dx.doi.org/10.4310/ATMP.2014.v18.n1.a2}{\emph{Adv. Theor. Math.
  Phys.} {\bf 18} (2014) 27--127}, [\href{http://arxiv.org/abs/1112.3984}{{\tt
  1112.3984}}].

\bibitem{Maruyoshi:2013fwa}
K.~Maruyoshi, C.~Y. Park and W.~Yan, \emph{{BPS spectrum of Argyres-Douglas
  theory via spectral network}},
  \href{http://dx.doi.org/10.1007/JHEP12(2013)092}{\emph{JHEP} {\bf 12} (2013)
  092}, [\href{http://arxiv.org/abs/1309.3050}{{\tt 1309.3050}}].

\bibitem{Beem:2013sza}
C.~Beem, M.~Lemos, P.~Liendo, W.~Peelaers, L.~Rastelli and B.~C. van Rees,
  \emph{{Infinite Chiral Symmetry in Four Dimensions}},
  \href{http://dx.doi.org/10.1007/s00220-014-2272-x}{\emph{Commun. Math. Phys.}
  {\bf 336} (2015) 1359--1433}, [\href{http://arxiv.org/abs/1312.5344}{{\tt
  1312.5344}}].

\bibitem{Liendo:2015ofa}
P.~Liendo, I.~Ramirez and J.~Seo, \emph{{Stress-tensor OPE in N=2
  Superconformal Theories}},  \href{http://arxiv.org/abs/1509.00033}{{\tt
  1509.00033}}.

\bibitem{Lemos:2015orc}
M.~Lemos and P.~Liendo, \emph{{$\mathcal{N}=2$ central charge bounds from $2d$
  chiral algebras}},
  \href{http://dx.doi.org/10.1007/JHEP04(2016)004}{\emph{JHEP} {\bf 04} (2016)
  004}, [\href{http://arxiv.org/abs/1511.07449}{{\tt 1511.07449}}].

\bibitem{Cordova:2015nma}
C.~Cordova and S.-H. Shao, \emph{{Schur Indices, BPS Particles, and
  Argyres-Douglas Theories}},
  \href{http://dx.doi.org/10.1007/JHEP01(2016)040}{\emph{JHEP} {\bf 01} (2016)
  040}, [\href{http://arxiv.org/abs/1506.00265}{{\tt 1506.00265}}].

\bibitem{Iqbal:2012xm}
A.~Iqbal and C.~Vafa, \emph{{BPS Degeneracies and Superconformal Index in
  Diverse Dimensions}},
  \href{http://dx.doi.org/10.1103/PhysRevD.90.105031}{\emph{Phys. Rev.} {\bf
  D90} (2014) 105031}, [\href{http://arxiv.org/abs/1210.3605}{{\tt
  1210.3605}}].

\bibitem{Cecotti:2015lab}
S.~Cecotti, J.~Song, C.~Vafa and W.~Yan, \emph{{Superconformal Index, BPS
  Monodromy and Chiral Algebras}},  \href{http://arxiv.org/abs/1511.01516}{{\tt
  1511.01516}}.

\bibitem{Cordova:2016uwk}
C.~Cordova, D.~Gaiotto and S.-H. Shao, \emph{{Infrared Computations of Defect
  Schur Indices}},  \href{http://arxiv.org/abs/1606.08429}{{\tt 1606.08429}}.

\bibitem{Buican:2015ina}
M.~Buican and T.~Nishinaka, \emph{{On the superconformal index of
  Argyres--Douglas theories}},
  \href{http://dx.doi.org/10.1088/1751-8113/49/1/015401}{\emph{J. Phys.} {\bf
  A49} (2016) 015401}, [\href{http://arxiv.org/abs/1505.05884}{{\tt
  1505.05884}}].

\bibitem{Buican:2015tda}
M.~Buican and T.~Nishinaka, \emph{{Argyres-Douglas Theories, the Macdonald
  Index, and an RG Inequality}},
  \href{http://dx.doi.org/10.1007/JHEP02(2016)159}{\emph{JHEP} {\bf 02} (2016)
  159}, [\href{http://arxiv.org/abs/1509.05402}{{\tt 1509.05402}}].

\bibitem{Song:2015wta}
J.~Song, \emph{{Superconformal indices of generalized Argyres-Douglas theories
  from 2d TQFT}}, \href{http://dx.doi.org/10.1007/JHEP02(2016)045}{\emph{JHEP}
  {\bf 02} (2016) 045}, [\href{http://arxiv.org/abs/1509.06730}{{\tt
  1509.06730}}].

\bibitem{Gadde:2009kb}
A.~Gadde, E.~Pomoni, L.~Rastelli and S.~S. Razamat, \emph{{S-duality and 2d
  Topological QFT}},
  \href{http://dx.doi.org/10.1007/JHEP03(2010)032}{\emph{JHEP} {\bf 1003}
  (2010) 032}, [\href{http://arxiv.org/abs/0910.2225}{{\tt 0910.2225}}].

\bibitem{Gadde:2010te}
A.~Gadde, L.~Rastelli, S.~S. Razamat and W.~Yan, \emph{{The Superconformal
  Index of the $E_6$ SCFT}},
  \href{http://dx.doi.org/10.1007/JHEP08(2010)107}{\emph{JHEP} {\bf 1008}
  (2010) 107}, [\href{http://arxiv.org/abs/1003.4244}{{\tt 1003.4244}}].

\bibitem{Gadde:2011ik}
A.~Gadde, L.~Rastelli, S.~S. Razamat and W.~Yan, \emph{{The 4d Superconformal
  Index from q-deformed 2d Yang-Mills}},
  \href{http://dx.doi.org/10.1103/PhysRevLett.106.241602}{\emph{Phys. Rev.
  Lett.} {\bf 106} (2011) 241602}, [\href{http://arxiv.org/abs/1104.3850}{{\tt
  1104.3850}}].

\bibitem{Gadde:2011uv}
A.~Gadde, L.~Rastelli, S.~S. Razamat and W.~Yan, \emph{{Gauge Theories and
  Macdonald Polynomials}},
  \href{http://dx.doi.org/10.1007/s00220-012-1607-8}{\emph{Commun.Math.Phys.}
  {\bf 319} (2013) 147--193}, [\href{http://arxiv.org/abs/1110.3740}{{\tt
  1110.3740}}].

\bibitem{Gaiotto:2012xa}
D.~Gaiotto, L.~Rastelli and S.~S. Razamat, \emph{{Bootstrapping the
  superconformal index with surface defects}},
  \href{http://dx.doi.org/10.1007/JHEP01(2013)022}{\emph{JHEP} {\bf 1301}
  (2013) 022}, [\href{http://arxiv.org/abs/1207.3577}{{\tt 1207.3577}}].

\bibitem{Minahan:1996fg}
J.~A. Minahan and D.~Nemeschansky, \emph{{An N=2 superconformal fixed point
  with E(6) global symmetry}},
  \href{http://dx.doi.org/10.1016/S0550-3213(96)00552-4}{\emph{Nucl. Phys.}
  {\bf B482} (1996) 142--152}, [\href{http://arxiv.org/abs/hep-th/9608047}{{\tt
  hep-th/9608047}}].

\bibitem{collingwood17008nilpotent}
D.~Collingwood and W.~McGovern, \emph{Nilpotent orbits in semisimple lie
  algebras van nostrand reinhold, new york 1993}, {\emph{Zbl0972} {\bf 17008}
  }.

\bibitem{Chacaltana:2011ze}
O.~Chacaltana and J.~Distler, \emph{{Tinkertoys for the $D_N$ series}},
  \href{http://dx.doi.org/10.1007/JHEP02(2013)110}{\emph{JHEP} {\bf 02} (2013)
  110}, [\href{http://arxiv.org/abs/1106.5410}{{\tt 1106.5410}}].

\bibitem{Chacaltana:2013oka}
O.~Chacaltana, J.~Distler and A.~Trimm, \emph{{Tinkertoys for the Twisted
  D-Series}}, \href{http://dx.doi.org/10.1007/JHEP04(2015)173}{\emph{JHEP} {\bf
  04} (2015) 173}, [\href{http://arxiv.org/abs/1309.2299}{{\tt 1309.2299}}].

\bibitem{Lemos:2012ph}
M.~Lemos, W.~Peelaers and L.~Rastelli, \emph{{The superconformal index of class
  $S$ theories of type $D$}},
  \href{http://dx.doi.org/10.1007/JHEP05(2014)120}{\emph{JHEP} {\bf 05} (2014)
  120}, [\href{http://arxiv.org/abs/1212.1271}{{\tt 1212.1271}}].

\bibitem{Panyushev201515}
D.~I. Panyushev, \emph{The dynkin index and $sl_2$-subalgebras of simple lie
  algebras},
  \href{http://dx.doi.org/http://dx.doi.org/10.1016/j.jalgebra.2015.01.033}{\emph{Journal
  of Algebra} {\bf 430} (2015) 15 -- 25}.

\bibitem{Shapere:2008un}
A.~D. Shapere and Y.~Tachikawa, \emph{{A Counterexample to the 'a-theorem'}},
  \href{http://dx.doi.org/10.1088/1126-6708/2008/12/020}{\emph{JHEP} {\bf 12}
  (2008) 020}, [\href{http://arxiv.org/abs/0809.3238}{{\tt 0809.3238}}].

\bibitem{Chacaltana:2014jba}
O.~Chacaltana, J.~Distler and A.~Trimm, \emph{{Tinkertoys for the E$_{6}$
  theory}}, \href{http://dx.doi.org/10.1007/JHEP09(2015)007}{\emph{JHEP} {\bf
  09} (2015) 007}, [\href{http://arxiv.org/abs/1403.4604}{{\tt 1403.4604}}].

\bibitem{Gadde:2015xta}
A.~Gadde, S.~S. Razamat and B.~Willett, \emph{{"Lagrangian" for a
  Non-Lagrangian Field Theory with $\mathcal N=2$ Supersymmetry}},
  \href{http://dx.doi.org/10.1103/PhysRevLett.115.171604}{\emph{Phys. Rev.
  Lett.} {\bf 115} (2015) 171604}, [\href{http://arxiv.org/abs/1505.05834}{{\tt
  1505.05834}}].

\bibitem{Kinney:2005ej}
J.~Kinney, J.~M. Maldacena, S.~Minwalla and S.~Raju, \emph{{An Index for 4
  dimensional super conformal theories}},
  \href{http://dx.doi.org/10.1007/s00220-007-0258-7}{\emph{Commun. Math. Phys.}
  {\bf 275} (2007) 209--254}, [\href{http://arxiv.org/abs/hep-th/0510251}{{\tt
  hep-th/0510251}}].

\bibitem{Romelsberger:2005eg}
C.~Romelsberger, \emph{{Counting chiral primaries in N = 1, d=4 superconformal
  field theories}},
  \href{http://dx.doi.org/10.1016/j.nuclphysb.2006.03.037}{\emph{Nucl. Phys.}
  {\bf B747} (2006) 329--353}, [\href{http://arxiv.org/abs/hep-th/0510060}{{\tt
  hep-th/0510060}}].

\bibitem{PANote}
P.~Agarwal. Unpublished.

\bibitem{Morita:2011cs}
T.~Morita and V.~Niarchos, \emph{{F-theorem, duality and SUSY breaking in
  one-adjoint Chern-Simons-Matter theories}},
  \href{http://dx.doi.org/10.1016/j.nuclphysb.2012.01.003}{\emph{Nucl. Phys.}
  {\bf B858} (2012) 84--116}, [\href{http://arxiv.org/abs/1108.4963}{{\tt
  1108.4963}}].

\bibitem{Agarwal:2012wd}
P.~Agarwal, A.~Amariti and M.~Siani, \emph{{Refined Checks and Exact Dualities
  in Three Dimensions}},
  \href{http://dx.doi.org/10.1007/JHEP10(2012)178}{\emph{JHEP} {\bf 10} (2012)
  178}, [\href{http://arxiv.org/abs/1205.6798}{{\tt 1205.6798}}].

\bibitem{Safdi:2012re}
B.~R. Safdi, I.~R. Klebanov and J.~Lee, \emph{{A Crack in the Conformal
  Window}}, \href{http://dx.doi.org/10.1007/JHEP04(2013)165}{\emph{JHEP} {\bf
  04} (2013) 165}, [\href{http://arxiv.org/abs/1212.4502}{{\tt 1212.4502}}].

\bibitem{Buican:2014hfa}
M.~Buican, S.~Giacomelli, T.~Nishinaka and C.~Papageorgakis,
  \emph{{Argyres-Douglas Theories and S-Duality}},
  \href{http://dx.doi.org/10.1007/JHEP02(2015)185}{\emph{JHEP} {\bf 02} (2015)
  185}, [\href{http://arxiv.org/abs/1411.6026}{{\tt 1411.6026}}].

\bibitem{DelZotto:2015rca}
M.~Del~Zotto, C.~Vafa and D.~Xie, \emph{{Geometric engineering, mirror symmetry
  and $ 6{\mathrm{d}}_{\left(1,0\right)}\to
  4{\mathrm{d}}_{\left(\mathcal{N}=2\right)} $}},
  \href{http://dx.doi.org/10.1007/JHEP11(2015)123}{\emph{JHEP} {\bf 11} (2015)
  123}, [\href{http://arxiv.org/abs/1504.08348}{{\tt 1504.08348}}].

\bibitem{Cecotti:2015hca}
S.~Cecotti and M.~Del~Zotto, \emph{{Higher S-dualities and Shephard-Todd
  groups}}, \href{http://dx.doi.org/10.1007/JHEP09(2015)035}{\emph{JHEP} {\bf
  09} (2015) 035}, [\href{http://arxiv.org/abs/1507.01799}{{\tt 1507.01799}}].

\bibitem{Xie:2016uqq}
D.~Xie and S.-T. Yau, \emph{{New N = 2 dualities}},
  \href{http://arxiv.org/abs/1602.03529}{{\tt 1602.03529}}.

\bibitem{Kutasov:1995ve}
D.~Kutasov, \emph{{A Comment on duality in N=1 supersymmetric nonAbelian gauge
  theories}}, \href{http://dx.doi.org/10.1016/0370-2693(95)00392-X}{\emph{Phys.
  Lett.} {\bf B351} (1995) 230--234},
  [\href{http://arxiv.org/abs/hep-th/9503086}{{\tt hep-th/9503086}}].

\bibitem{Kutasov:1995np}
D.~Kutasov and A.~Schwimmer, \emph{{On duality in supersymmetric Yang-Mills
  theory}}, \href{http://dx.doi.org/10.1016/0370-2693(95)00676-C}{\emph{Phys.
  Lett.} {\bf B354} (1995) 315--321},
  [\href{http://arxiv.org/abs/hep-th/9505004}{{\tt hep-th/9505004}}].

\bibitem{Dolan:2008qi}
F.~A. Dolan and H.~Osborn, \emph{{Applications of the Superconformal Index for
  Protected Operators and q-Hypergeometric Identities to N=1 Dual Theories}},
  \href{http://dx.doi.org/10.1016/j.nuclphysb.2009.01.028}{\emph{Nucl. Phys.}
  {\bf B818} (2009) 137--178}, [\href{http://arxiv.org/abs/0801.4947}{{\tt
  0801.4947}}].

\bibitem{Spiridonov:2009za}
V.~P. Spiridonov and G.~S. Vartanov, \emph{{Elliptic Hypergeometry of
  Supersymmetric Dualities}},
  \href{http://dx.doi.org/10.1007/s00220-011-1218-9}{\emph{Commun. Math. Phys.}
  {\bf 304} (2011) 797--874}, [\href{http://arxiv.org/abs/0910.5944}{{\tt
  0910.5944}}].

\end{thebibliography}\endgroup

\end{document}